# Mid-Infrared Spectroscopy of Uranus from the Spitzer Infrared Spectrometer: 1. Determination of the Mean Temperature Structure of the Upper Troposphere and Stratosphere

Glenn S. Orton[a], Leigh N. Fletcher[b], J. I. Moses[c], A. K. Mainzer[d], Dean Hines[e], H. B. Hammel[f], F. J. Martin-Torres[g], Martin Burgdorf[h], Cecile Merlet[b], Michael R. Line[i]

[a]MS 183-501, Jet Propulsion Laboratory, California Institute of Technology, 4800 Oak Grove Drive, Pasadena, California 91109, USA

[b]Atmospheric, Oceanic & Planetary Physics, Clarendon Laboratory, University of Oxford, Parks Road, Oxford OX1 3PU, UK

[c]Space Science Institute, 4750 Walnut St., Suite 205, Boulder, CO 80301, USA

[d]MS 321-535, Jet Propulsion Laboratory, California Institute of Technology, 4800 Oak Grove Drive, Pasadena, California 91109, USA

[e]Space Telescope Science Institute, 3700 San Martin Drive, Baltimore, MD 21218, USA

[f]Association of Universities for Research in Astronomy, 1212 New York Avenue NW, Suite 450, Washington, DC 20005, USA

[g]Instituto Andaluz de Ciencias de la Tierra (CSIC-INTA), Avda. De las Palmeras, 4, 18100, Armilla, Granada, Spain

[h]HE Space Operations, Flughafenallee 24, D-28199 Bremen, Germany

[i]Department of Astronomy and Astrophysics, University of California-Santa Cruz, Santa Cruz, California 95064, USA

Corresponding author: G. Orton[a]
glenn.orton@jpl.nasa.gov
818-354-2460

              Infrared observations
              Atmospheres, structure

ABSTRACT

On 2007 December 16-17, spectra were acquired of the disk of Uranus by the Spitzer Infrared Spectrometer (IRS), ten days after the planet's equinox, when its equator was close to the sub-earth point. This spectrum provides the highest-resolution broad-band spectrum ever obtained for Uranus from space, allowing a determination of the disk-averaged temperature and molecule composition to a greater degree of accuracy than ever before. The temperature profiles derived from the Voyager radio occultation experiment by Lindal et al. (1987, J. Geophys. Res. 92, 14987) and revisions suggested by Sromovsky et al. (2011, Icarus 215, 292) that match these data best are those that assume a high abundance of methane in the deep atmosphere. However, none of these model profiles provides a satisfactory fit over the full spectral range sampled. This result could be the result of spatial differences between global and low-latitudinal regions, changes in time, missing continuum opacity sources such as stratospheric hazes or unknown tropospheric constituents, or undiagnosed systematic problems with either the Voyager radio-occultation or the Spitzer IRS data sets. The spectrum is compatible with the stratospheric temperatures derived from the Voyager ultraviolet occultations measurements by Herbert et al. (1987. J. Geophys. Res. 92, 15093), but it is incompatible with the hot stratospheric temperatures derived from the same data by Stevens et al. (1993, Icarus 100, 45). Thermospheric temperatures determined from the analysis of the observed $H_2$ quadrupole emission features are colder than those derived by Herbert et al. at pressures less than ~1 microbar. Extrapolation of the nominal model spectrum to far-infrared through millimeter wavelengths shows that the spectrum arising solely from $H_2$ collision-induced absorption is too warm to reproduce observations between wavelengths of 0.8 and 3.3 mm. Adding an additional absorber such as $H_2S$ provides a reasonable match to the spectrum, although a unique identification of the responsible absorber is not yet possible with available data. An immediate practical use for the spectrum resulting from this model is to establish a high-precision continuum flux model for use as an absolute radiometric standard for future astronomical observations.

# 1. Introduction

The mid-infrared spectra of planets are replete with information on their composition and temperature structure. The spectrum of Uranus is particularly challenging because of its faintness. Nevertheless, thermal infrared spectra of Uranus at 7.7 μm and longer wavelengths provide opportunities to sense temperatures over a wide vertical range stretching from as low as ~1 nanobar up to 2 bars of atmospheric pressure (Orton et al. 1987, Orton et al. 2007, Werner et al. 2004, and Houck et al. 2004). At wavelengths shorter than 7.7 μm, scattered sunlight dominates the spectrum. At longer wavelengths, much of the spectrum is dominated by the opacity of $H_2$ collision-induced absorption (CIA). Emission features are also detectable from strong quadrupole lines of molecular hydrogen ($H_2$), the $CH_4$ band at 7.7 μm, and for the hydrocarbon photolysis products acetylene ($C_2H_2$) at 13.7 μm, ethane ($C_2H_6$) at 12.2 μm, methylacetylene ($CH_3C_2H$) at 15.8 μm, and diacetylene ($C_4H_2$) at 15.9 μm (Burgdorf et al. 2006).

Because of its extreme orbital obliquity of 98°, Uranus undergoes severe changes in seasonal insolation that could affect its atmospheric temperature, composition and structure. Every 42 years at the equinox, both the northern and southern hemispheres may be viewed from the Earth simultaneously. The Spitzer Space Telescope Infrared Spectrograph (IRS) had been used to acquire disk-averaged spectra of Uranus from 5 to 37 μm in Spitzer Guaranteed Time Observations (GTO) in 2004 (Burgdorf et al. 2006), and in 2005 during Spitzer's Cycle 1 (Orton et al. 2007). We report here observations made shortly after Uranus' 2007 December 7 equinox. These observations were motivated by the opportunity to detect any variability of the disk-averaged spectrum that might result from winter pole coming into view or any genuine temporal changes of the atmosphere. We used our experience with the earlier Spitzer IRS observations of Uranus to improve our approach and create the most sensitive spectral measurements of the planet in the mid-infrared spectral region.

In this paper we describe our approach to deriving the globally averaged vertical temperature structure of Uranus. Our procedure is based on matching spectral regions dominated by the continuum-like $H_2$ CIA, together with its S(1), S(2) and S(3) quadrupole lines, because $H_2$ is well mixed throughout the atmosphere and its molar fraction only change significantly in regions where a major constituent such as $CH_4$ condenses, a factor that is accounted for in our models (see Paper 2). The sensitivity of these spectra allows us access to the CIA signature in the 10-μm region -which is too faint for ground-based observations (e.g., Orton et al. 1987, 1990) - and thus to sense temperatures as deep as 2 bars of atmospheric pressure, as illustrated by Orton et al. (2007) with earlier IRS spectra. These depths were unavailable to the Voyager IRIS experiment (see Flasar et al. 1987, Conrath et al. 1998).

We then compare our results with previous data and models for Uranus. These include the Voyager-2 results from the IRIS experiment (Flasar et al. 1987, Conrath et al. 1998) and the radio-occultation experiment (Lindal et al. 1987), including the re-interpretation of radio-occultation results by Sromovsky et al. (2011) and their implications for the bulk composition of Uranus. We examine the implications of our nominal model for the far-infrared through millimeter spectrum and its utility to the Herschel mission and other missions and investigations as a calibration standard. Finally, we summarize our results and suggest future clarifying work. A companion paper (Orton et al. 2014) discusses the analysis of these spectra for the global-mean composition of Uranus, hereafter referenced as Paper 2.

## 2. Spitzer Uranus data

**2.1 Acquisition and spectral extraction**. Following successful observations of Uranus on 2005 July 6-7 (Orton et al. 2007), we obtained an additional 9.2 hours of Director's Discretionary Time (program #467) to document changes in the planet's atmosphere. The observations were performed on 2007 December 16-17, just ten days after Uranus' equinox. We used the Spitzer IRS, a high-resolution grating spectrometer with no moving parts and four different modules, Short High (SH), Short Low (SL), Long High (LH) and Long Low (LL). All four modules were used, covering the wavelength range of 5-37 μm, with spectral resolutions $R = \Delta\lambda/\lambda = 90 - 600$. Because only a single IRS module can be used at one time, each longitude observation was necessarily taken at a different time.

**Table 1** summarizes the characteristics of the different modules, and **Table 2** summarizes the observations. Four epochs were equally spaced over one equatorial rotation period of the planet (17.25 hours). The angular diameter of Uranus was 3.35 arcsec from the vantage point of Spitzer, and the Spitzer beam size ranged from 3.6 to 10 arcsec going from 5 to 37 μm (**Table 1**). Because the disk of Uranus was never resolved to better than 50%, we extracted only a flux at each longitude that is averaged over the slit. **Table 3** gives the time at the start of each module observation and the longitude to which it corresponds for a rotational rate coupled to the magnetic field.

>Table 1.

We conducted separate observations of the off-source sky with both the SH and LH modules (see **Table 2**). The purpose of the off-source observations was twofold: (1) the zodiacal background is non-negligible at mid-infrared wavelengths and (2) radiation had permanently damaged some of the instrument's pixels, resulting in high dark currents. Contemporaneous off-source observations allowed most of the dark current to be subtracted; the remainder can be readily identified in the off-source images.

>Table 2.

Bad pixels are more prominent in the high-resolution modules due to the smaller number of photons per wavelength bin. The effects of the bad pixels were mitigated by using the Spitzer Science Center's (SSC) recommended algorithm, *irsclean_mask*, using campaign-specific bad pixel masks supplied by the SSC. Any remaining pixels that were not correctly identified by the SSC routine were found by visually inspecting the two-dimensional images and adding them to the SSC-supplied bad pixel masks, then rerunning *irsclean_mask*.

Spitzer observes with IRS by nodding between two positions located one-third and two-thirds along the slit of each module. The images for both nods were coadded into single frames, and the appropriate sky observations were subtracted. For the SH and LH modules, the off-source pointings were used. However, no specific off-source pointings were required for the low-resolution modules. Both SL and LL modules are each divided into two separate sub-modules covering different spectral orders. The SL module is divided into SL1 and SL2, each requiring a unique telescope pointing; therefore, while the source is in one order, the other order is observing blank sky. Additionally, both SL and LL modules have slit lengths much longer than the size of the image (57 and 128 arcsec, respectively), allowing the sky to be extracted directly from the image. Both sky-subtraction techniques are used for SL1 and SL2. No LL1 data were acquired because of expected detector saturation in these long wavelengths. Because we only observed with LL2, we subtract the sky directly from the image at the opposite end of the slit from the source.

All spectra are extracted using the SSC-provided Spitzer IRS Custom Extraction (SPICE) tool using the standard point-source extraction template. The high-resolution modules were extracted using the full slit width, and the l00-resolution modules using a tapered column extraction window. The LH data are affected by fringes whose frequency is a function of the thickness of the detector substrate and have to be corrected ("defringed") using the standard *irs_fringe* module from the SSC.

Data are converted to radiance based on a distance of 23.23 A.U. from Spitzer on December 16, 2007, with a sub-spacecraft latitude of 2.7°S and equatorial and polar radii of 25559 km and 24794 km, respectively (Lindal et al. 1987). The combined spectra are shown in **Figure 1**.

>Figure 1

Care was also taken in the calibration to account for overfilling of the slit by the disk of Uranus, convolved with the point-spread function for the IRS. This was calculated using the IRS tool *stinytim*, downloaded from the SSC's "Documentation and Tools" web site. This code generated point-spread functions for the IRS as a function of wavelength. We convolved these point-spread functions with ellipsoids equal to Uranus in size to simulate the appearance of Uranus to the IRS as a function of wavelength and thus to determine the extent of flux lost from the slit, accounting for the 29.41° clockwise rotation of the slit relative to the rotation axis of Uranus. Because the spectrum contained both absorption and emission features, we used a range of limb-darkened to limb-brightened models for the flux within each image, patterned after the most limb-darkened and limb-brightened of our models described later in this paper and Paper 2. A "limb-flat" assumption served as an intermediate case. **Figure 2A** demonstrates the favorable comparison of this model convolution of the point-spread function with a scan along the slit through the measured position of Uranus along a column of the detector passing closest to the disk center. For the 10.5"-wide LL slit, the corrections were on the order of 0.4% or less, much lower than our 3% measurement uncertainty, and we made no corrections to the measured flux. For the 3.6-3.7" tapered SL slit, 5-10% wavelength-dependent corrections were required, as shown in **Figure 2B** for the SL wavelengths of interest to this study, assuming a flat center-to-limb behavior. Uncertainties arising from model-dependent center-to-limb functions ranged from 2.5 to 3.8%. Rather than applying such model-dependent corrections, we simply assumed a conservative 4% value for this source of uncertainty. This might be reduced in the future using verification of the model via observations of the center-to-limb dependence. Although, for the sake of brevity, we will refer to all the reduced spectra shown in Figure 1 as disk-averaged, the SL spectra represent our best estimate of disk-averaged spectra and have an additional corresponding uncertainty.

>Figure 2

**2.2 Absolute calibration:** A thorough examination of the uncertainties associated with the Spitzer calibration system has been made recently by Sloan et al. (2011). Their discussion of the absolute calibration system notes two sources of uncertainty. One is a disagreement on what should be assumed as standard stars, with a difference of roughly 4%. The other arises from differences in the infrared calibrations by, for example, Rieke et al. (2008) and Engelke et al. (2010), which adds another 2%. Combining these with our 3% measurement uncertainty (see section 2.3.1) yields a total flux or radiance uncertainty that is nearly 6% for the LL spectra. For the SL spectra, we must add an additional 4% uncertainty arising from the slit overfilling correction, for a cumulative uncertainty of 7%. As described below, the SL and LL spectra are considered photometrically accurate to these levels, but the SH and LH slits are not and must be corrected by scaling to the low-resolution observations in spectral regions of overlap.

We do not cross-compare calibration with previous observations from the ground (Orton et al. 1987, 1990) or from the Infrared Space Observatory (ISO) Short Wavelength Spectrometer (SWS), e.g. Fouchet et al. (1990). First, those observations of Uranus represent very different aspect geometries, i.e. closer to pole-on and with a possibly very different polar temperature structure than our equator-on observations. Furthermore, the ground-based observations (Orton et al. 1987, 1990) have substantial uncertainty in absolute calibration that essentially precludes meaningful conclusions. The ISO SWS observations also suffer from a substantial uncertainty of 0.5-1.0 Jy in the dark-current level (Leech et al. 2003; H. Feuchtgruber, pers. comm.), which corresponds to an offset of at least 45-90% of the observed flux in the spectral region we report here.

On the other hand, a meaningful cross-check of the calibration is possible using an observation of Uranus by the broad 6.2-9.2-µm filter of the Spitzer IRAC instrument that was made on 2006 June 12. The observation is independent both in the sense that it was made using a different instrument and in that it used a different stellar calibration system. Convolving the spectral function of the IRAC filter over our SL2 spectrum yields a value that is only 5% higher than the measured IRAC radiance. This difference is well within the 7% uncertainty associated with SL spectra.

**2.3 Consistency of calibrated spectra**. Prior to detailed modeling of the spectra, the low-resolution and high-resolution modes are co-plotted to assess the consistency of the calibration. We examine three regions shown in **Table 2**: (a) the low-resolution (LL2, SL1) and high-resolution (SH, LH) modules overlap in the 14.0-19.6 µm region, sensitive to the wings of the $C_2H_2$ feature and the collision-induced $H_2$ continuum; (b) the SL1 and SH overlap in the 9.9-14.5 µm region, sensitive to hydrocarbon emission features; and (c) the SL1 and SL2 module overlap in the 7.4-7.7 µm region, sensitive to $CH_4$ emission.

**2.3.1 The 14.0-19.6 µm region**: The low-resolution modules (SL and LL) yield the most self-consistent results. For each module, we estimate the uncertainty arising solely from the reproducibility of the flux measurements to be 3%. However, in comparing spectra from the SL and LL modules in their overlap region (**Fig. 3**), an inconsistency is apparent at the location of the acetylene emission (averaged over all four longitudes). The LL2 flux is 9% higher than the SL1 flux in their area of overlap. Although one might attribute this behavior to spillover in the SL1 slit compared with the wider LL2 slit, we correct for this in some detail. The discrepancy would be approximately 18% without the correction. We also checked that Uranus was centered in the slit for each spectrum. In fact, the discrepancy is roughly commensurate with both (i) differences seen in the shorter-wavelength orders described below and (ii) a commensurate range of radiance offsets of individual SL1 and LL2 spectra of hydrocarbon emission described below in section 2.5. The difference is thus far more likely to be the result of rotational variability of hydrocarbon emission that is sampled at different longitudes (Table 3), particularly because the $H_2$-dominated regions of the SL1 and LL2 spectra are independently reproducible to within the 3% measurement uncertainty.

>Figure 3
>Figure 4
>Figure 5

**Figure 4** compares the LL2 mean spectrum with SH and LH overlap regions around 14-22 µm. LL2 nod positions and longitudes are consistent with each other throughout the $H_2$ continuum region, but begin to show deviations at the 10% level as we approach the acetylene feature at the short-wavelength end of the module. No part of the LL2 spectrum is saturated, so the entire range is deemed reliable. Longitudes 1 and 4 are consistently warmer than longitudes 2 and 3 in the acetylene feature. Because of the consistency of the LL2 spectra in the continuum, we scale the LH and SH spectra to match the LL2 and SL1 fluxes. A multiplicative scaling, varying between individual orders but with a mean of 12%, is used to make an initial correction to the SH and LH spectra. In addition to the scaling, a spectral slope was detected in the SH and LH spectra, relative to the SL1 and LL2 fluxes, which also required a "pivoting" correction, as illustrated in **Figure 5**. An offset, rather than a scaling, would also provide a good initial fit, but subsequent "pivoting" failed to match the low-resolution spectra as well as the scaling + pivoting approach. The uncertainties associated with scaling of SH spectral orders to either SL1 or LL2 fluxes are extremely low, on the order of 1-2% or less.

**2.3.2 The 9.9-14.5 µm region**: The short-wavelength orders of the SH module are inconsistent, both internally between orders and with the SL1 spectra, sometimes by factors of several. **Figure 6** shows the overlap between the high- and low-resolution modes, demonstrating the difficulties associated with the short-wavelength end of the SH module. This figure also breaks down the mean SH spectrum into its separate orders (coadded for each longitude and nod), and demonstrates that inconsistencies arise between the individual orders, becoming worse at shorter wavelengths. As a result, our analysis omits SH orders at wavelengths shorter than 11.9 μm, and care must be taken with the radiometric accuracy of the SH mode.

>Figure 6

The same is true of the LH mode. Although the high signal of the LH module (Fig. 1) makes the internal inconsistencies appear negligible, after much effort we were unable to find physically meaningful models fitting the LH spectrum. Because there is no low-resolution coverage of much of the wavelength range for the LH mode on which its absolute flux calibration could be based, we deem this portion of the spectrum unreliable and are thus forced to ignore the entire 19.6-37.2 μm LH spectrum in our analysis.

**2.3.3 The 7.4-8.7 μm region**: Finally we look at the overlap of the SL1 and SL2 spectra, along with the "SL bonus order" which overlaps with both modes (covering 7.3-8.6 μm with the same spectral resolution as SL1). Although SL2 and the bonus order are taken at the same time, SL1 requires a separate pointing of the telescope and so is separated in time. **Figure 7** demonstrates potential inconsistencies between the observations made in the SL1, "bonus", and SL2 orders. The average SL2 and "bonus" spectra differ from each other by only 6% in overlapping wavelengths. The average SL1 and SL2 spectra differ by 20%; however, they did not cover the same longitudes (**Table 3**), and different longitudes exhibit different methane emission intensities. The global-average conditions derived from coadded spectra must therefore have some degree of spatial uncertainty.

>Figure 7
>Table 3

**2.4 Spectral resolution and wavelength uncertainties**: The dependence of the spectral resolution on wavelength for the high-resolution modules varies between R ($\lambda/\Delta\lambda$) ≈ 540 – 660 according to the IRS Observers Handbook (http://irsa.ipac.caltech.edu/data/SPITZER/docs/irs/irsinstrumenthandbook/). These values, calibrated from well-known galaxies or planetary nebula, were only measured at a few discrete wavelengths. Our analyses of the Uranus spectra indicates that spectral resolutions vary over a somewhat wider range, as low as 611 and as high as 743. The cause of the wavelength-dependent variation of spectral resolution remains unknown, and because it is not well documented for every order, it was treated as a free parameter with a uniform value over each order. At the longest wavelengths, diffraction smears the different wavelengths over two pixels, in any case. Changes of the instrument function - e.g., triangular or flat ("boxcar") rather than Gaussian - yield insignificant changes in solutions. We note that Doppler smearing from planetary rotation is detectable only a resolving powers great than $10^5$ and is thus several orders of magnitude smaller than our spectral resolution in any of the modules. The wavelength calibration of IRS is good to between 20-25% of a resolution element. We note that we must apply a constant wavelength offset of +0.08 cm$^{-1}$ to the SH orders to allow accurate fitting of the S(1) quadrupole. No such requirement is necessary for the S(2) quadrupole transition nor the lines of ethane surrounding it.

**2.5 Variability over time**. The December 2007 equinox dataset reveals surprising rotational variability in the strength of the hydrocarbon emissions. The variability is consistent with a longitudinal dependence of the hydrocarbon emissions that will be the topic of a subsequent paper. Interestingly, neither the $H_2$ continuum nor the $H_2$ quadrupole varies as a function of longitude above the expected 3% level of measurement uncertainty.

Over a time scale of years, we note few substantial differences between these observations and similar spectra taken in 2005, as shown in **Figure 8**, which displays the SL spectra from each of these epochs. Significant differences only appear at wavelengths less than 7 μm, which is increasingly affected by a contribution from reflected sunlight. Our analysis addresses only observations made at wavelengths longer than 7 μm, where the contribution of reflected sunlight is expected to be negligible, as are differences between these and the Cycle-1 observations relative to the noise level. We concentrate on the 2007 observations here because of their superior signal-to-noise ratios and the presence of LL spectra, which were not made in our 2005 observations.

>Figure 8.

**3. Radiative transfer**

The calculations of upwelling radiance described in the following sections are made with a 10-stream trapezoidal quadrature in the emission angle cosine to determine a disk-averaged value of the upwelling radiance. The values of the disk-averaged radiance are adjusted to a planetary radius equal to our 100-nanobar minimum pressure level, which is sufficiently high in the atmosphere to integrate over the full range of contributions for the outgoing radiance at all wavelengths, except for $H_2$ quadrupole line emission. In those cases, the minimum pressure level is lowered to 1 nanobar. These considerations limit errors in determining the disk-averaged radiance from numerical quadrature to less than 0.5% in the worst case for the calculations that involved pressures greater than or equal to 100 nanobars. For calculations involving quadrupole lines that involve pressures as low as a nanobar, the worst case is 5%, the consequence of which is an uncertainty of 1K or less in constraining upper stratospheric and thermospheric temperatures. Use of a more standard Gauss-Legendre quadrature renders the results much more sensitive to the choice of the lower-pressure boundary that represented the planetary limb because of the density of grid points near an emission-angle cosine of 0; this is not the case with the trapezoidal quadrature approach. Atmospheric parameters are specified, as required, to pressures that reached a maximum of 10 bars and a minimum of 1 nanobar.

**3.1 Spectroscopic line parameters**. We model $H_2$-$H_2$ collision-induced absorption (CIA) using the values tabulated by Orton et al. (2007) from *ab initio* models. $H_2$-He and $H_2$-$CH_4$ CIA value are modeled using similar tables, created from programs emulating the *ab initio* calculations of Borysow et al. (1988) and Borysow and Frommhold (1986), respectively. We add to the $H_2$-$H_2$ collision-induced absorption tables the contributions of $(H_2)_2$ dimers (Schaefer and McKellar 1990). Although we do not match discrete dimer features in this paper, we note their influence extends several cm$^{-1}$ away from line center, for example, affecting the continuum $H_2$-$H_2$ absorption around the S(1) 587 cm$^{-1}$ CIA rotation at the low temperatures of Uranus' atmosphere, as will be illustrated in Section 4.3.1.

Spectroscopic parameters for the $H_2$ quadrupole S(1), S(2), S(3) and S(4) lines are updated from those used in the analysis of ISO SWS data by Fouchet et al. (2003), namely Jennings et al. (1987) and Poll and Wolniewicz (1978). The update was reported by Campargue et al. (2012); an error in the intensities listed in their Table 3 was corrected in the version of line parameters submitted to the HITRAN2012 data base (Rothman et al. 2013). Line-broadening widths are taken from the work of Reuter and Sirota (1994).

We will describe spectroscopic parameters relevant to all the features observed in the spectrum in Paper 2. In this paper, we discuss $H_2$ quadrupole lines that are embedded among lines of $CH_4$ and $C_2H_6$. We use $CH_4$ and $CH_3D$ spectroscopic parameters as described by Brown et al. (2003), with line transitions, strengths, and quantum identifications for $CH_4$ taken from Ouardi et al. (1996) and for $CH_3D$ taken from Nikitin et al. (2000). $H_2$- and He-broadening widths were taken from Pine (1992) and Margolis (1996). Spectroscopic parameters for $C_2H_6$ are taken from Vander Auwera et al. (2007) using the version of their list that includes the $H_2$-broadening widths of Halsey et al. (1988); this list includes isotopic $^{13}C^{12}CH_6$ lines.

**3.2 Non-Local thermodynamic equilibrium**. For the $H_2$ quadrupole lines, local thermodynamic equilibrium (LTE) is expected to be valid at extremely low pressures (Trafton 1999). Thus, we assume LTE up to the nanobar level, where only insignificant changes to upwelling radiance at the centers of the **$H_2$** quadrupole lines take place. For all other lines, we use a non-LTE model that is an updated version of the methane model by Martín-Torres et al (1998) with a similar collisional scheme, which is explained in detail in Paper 2.

**3.3 Clear-atmosphere assumption.** In the early studies of this spectral region (Orton et al. 1987, 1990) it was necessary to invoke the opacity of stratospheric particles to explain the high radiances emerging from this region. The optical thickness of such a layer could be rather low because radiation from the cold upper troposphere of Uranus passes through a much warmer stratosphere. However, the substantial revision of the $H_2$ collision-induced opacity (Orton et al. 2007) obviated the need for this aerosol opacity, and so we omit such an assumption from our initial approach. In Section 5.4, we discuss in detail the properties of particulates in the warm stratosphere that would make significant contributions to the upwelling thermal radiance in the context of understanding differences between our analysis and Voyager radio-occultation results.

## 4. Deriving a global average temperature profile

Despite the stratospheric variability observed in the four longitudes measured in December 2007, we coadd all spectra to produce one 'mean' spectrum for each module. In this section, we derive an averaged temperature structure from the SH, LL2, and SL1 modules using an atmospheric model containing $H_2$, He and $CH_4$. The consistency of the spectra to within the 3% overall observational uncertainties for the $H_2$ continuum, and even the S(1) quadrupole feature, further supports this approach as being an accurate representation of average tropospheric and lower stratospheric temperatures. On the other hand, our further analysis of absorption and emission from other quadrupole lines, together with other stratospheric constituents must be taken as a representation of zonal-mean conditions that, in fact, had significant zonal variability. We intend to undertake the quantitative analysis of this variability in subsequent publications, reporting here only our best available estimate for global-mean conditions.

**4.1 First Approximation to the Uranian atmospheric structure.** We begin with the Voyager Radio SubSystem (RSS) occultation profile preferred by Lindal et al. (1987) as a first approximation, i.e. their model D. This profile extends from 2.3 bars to 2.5 mbar. If we use their Model F or Models EF or F1 of the same data analyzed by Sromovsky et al. (2011) as an initial profile, there are no differences in the results. The radio-occultation profile features high-altitude oscillations, which could be representative of either vertically propagating waves in the stratosphere or levels of localized radiative heating from hydrocarbon condensate haze layers that are not time-dependent oscillations. Because the vertical extent of the oscillations is below the resolution of our retrieval and they sample an extremely limited region of the planet, we substitute a straight line in the region of the stratospheric thermal oscillations, and extended it upward by assuming a constant temperature gradient, as shown in **Figure 9**. We vary the value of this lapse rate as a boundary condition in fitting the spectrum by retrieving the temperature profile at pressures greater than 2.5 mbar. As will be described in detail below, we find that it was also necessary to account for contributions from very high temperatures in the thermosphere of Uranus, at atmospheric pressures as low as 1 nanobar, in order to model the contributions from $H_2$ S(2) and S(3) quadrupoles. An initial temperature profile is taken that mimics profile 'b' in Fig. 7 of Herbert et al. (1987), their "compromise" profile that matches best both Voyager UVS experiment stellar and solar occultation data; this initial profile is shown in **Fig. 9**. For the purposes of modeling the continuum arising from deeper in the troposphere, the radio occultation profile is extrapolated down to 10 bars of atmospheric pressure, using an adiabatic lapse rate from the 2.3-bar level, assuming equilibrium mixtures of *para-* and *ortho-*$H_2$ at the local temperature without active heat exchange. This 'extended' Voyager profile served as our *a priori* temperature profile for fine-scale perturbation to fit the Spitzer spectra.

>Figure 9

The IRS spectra of Uranus do not provide the means for independent determination of the bulk composition, and we must rely on additional constraints. Following the Voyager IRIS analysis of Conrath et al. (1987), we initially assume a ratio of 85% $H_2$ to 15% He by number. To this mixture, a $CH_4$ mole fraction of 3.2% is added in the deeper atmosphere, below the methane condensation level, consistent with the results of Karkoschka and Tomasko (2009, 2011). This value is greater than the 2.3% mole fraction associated with the Voyager RSS profile of "model D" preferred by Lindal et al. (1987) but less than values of 3.8-4% in models matching the same data considered by Sromovsky et al. (2011). This deep abundance smoothly decreases with altitude up to the tropopause, above which the value of the $CH_4$ mole fraction remains constant with altitude. The influence of $H_2$-$CH_4$ collision-induced absorption on the 540-650 $cm^{-1}$ region of the spectrum was found to be significant, such that simple $H_2$- and $H_2$-He-only atmospheres failed to reproduce the Spitzer spectra. As described in Paper 2, it is necessary to vary the fraction of saturation ("humidity") of $CH_4$ at all altitudes above its condensation level in order to match its emission and absorption spectrum. Our tests of the effects of the adopted $CH_4$ mole fraction on the $H_2$-$CH_4$ collision-induced absorption demonstrate that $CH_4$ mole-fraction uncertainties do not have a significant effect on derived temperatures.

**4.2 Contribution functions.** We define the contribution functions as the product of the transmission weighting function $d\tau/dz$ and the black-body function $B(z,T)$, and indicate the altitudes to which each region of the spectrum is sensitive (which in turn depend on the temperature profile and composition). To guide our perturbations to the reference atmosphere, **Figure 10A** demonstrates the vertical sensitivity of the continuum regions of the spectrum of Uranus, together with the center of the $H_2$ S(1) line. In the 15.6-18.5 μm (548-640 $cm^{-1}$) region shown in **Fig. 10A**, most of the sensitivity is around the 50-400 mbar region. Some sensitivity to 1 mbar occurs over the 16.7-17.9 μm (560-600 $cm^{-1}$) region. Because the temperatures are so low, emission from the vicinity of the temperature minimum is negligible. As a result, for the continuum at 17.0 μm (587 $cm^{-1}$: the peak of the broad S(1) rotational line) the contribution is multi-lobed with peaks at 2 mbar and 100 mbar. The 2-mbar stratospheric contribution is only 25% of the tropospheric contribution.

>Figure 10

We resolve this degeneracy by initially matching observations of the $H_2$ quadrupoles by variations of our model temperature profile. We initially vary stratospheric temperatures in the 0.02-2.0 mbar range to fit both the continuum and the $H_2$ S(1) quadrupole. The contribution functions of the S(2) and S(3) quadrupoles are sensitive both to stratospheric temperatures and to temperatures at pressures below the 10-μbar level (**Fig. 10B**); we use them to determine both the pressure at which the rapid temperature increase takes place and the atmosphere's peak temperature. The 9.00-11.36 μm (880-1112 $cm^{-1}$) region probes even deeper, to pressures greater than 1 bar. At 11.36 μm (880 $cm^{-1}$), the contribution function peaks at 1 bar, with a width from 800-1400 mbar. The peak contribution in this range occurs at 9.1 μm (1100 $cm^{-1}$), where the peak absorption is at 2 bar, with a range between 1.2-3.0 bar.

**4.3 Deriving the temperature profile**. Based on the contribution function plots, we undertake a series of logical steps to derive the temperature profile. In the following text, the reader should bear in mind that the retrieval of temperatures is an inherently non-unique and underconstrained problem, with a potentially large number of solutions providing adequate fits to the data. To a large extent, our solutions attempt to produce a temperature profile that is the simplest possible one among several alternatives.

>Figure 11

**4.3.2. Perturb the lower stratosphere.** Standard temperature-sounding approaches (e.g. Chahine 1970, Rodgers 2000) work well for the deep atmosphere, but exert little influence over the low-temperature region around the tropopause. Using the 10-mbar level as a pivot point, we use an exponential function above and below this level to cool the lower stratosphere and make the tropopause region more isothermal. The pressure and height grid and the $CH_4$ saturation curve are recalculated for each modification. By varying the size of the perturbation (**Fig. 11A**), we arrive at a profile that produces a reasonable fit to the 18.2-19.2 μm (520-550 $cm^{-1}$) continuum (**Fig. 11B**).

**4.3.3. Tropospheric temperatures.** Starting from the lower-stratospheric perturbation from the previous step, we then use standard inversion approaches to derive temperatures in the deeper atmosphere from the $H_2$ CIA continuum at all wavelengths between 9.5 and 20.2 μm (447-1058 $cm^{-1}$) where it is not confused by spectral features arising from other molecules (see **Fig. 12B**). An adiabatic lapse rate is assumed below the 2-bar pressure level.

**4.3.4. Mid-stratospheric temperatures from the 17.04-μm (587 $cm^{-1}$) SH spectrum.** Along with these perturbations, we also test a range of values for the overlying stratospheric lapse rate, as shown in **Fig. 12A**. For each lapse rate tested, we vary the lower-stratospheric perturbation described in 4.3.2 and derive temperatures using a formal retrieval as described in 4.3.3. An optimal fit that minimizes residuals has a strong temperature inversion, as shown by the solid line in **Figure 12A**. The sensitivity of the $H_2$ collision-induced spectral continuum to these solutions is shown in **Figures 12B** and **12C** and the $H_2$ S(1) quadrupole at 587 $cm^{-1}$ in **Figures 12D, 13C and 13D**.

>Figure 12

**4.3.5. Matching the S(2), S(3) and S(4) $H_2$ quadrupole lines.** Matching the 12.3-μm (814 $cm^{-1}$) $H_2$ S(2) line is complicated by its appearance in the middle of $C_2H_6$ emission, but a fit to the surrounding emission is well approximated by scaling the vertical distribution of $C_2H_6$ from an update of the Moses et al. (2005) photochemical model described in Paper 2 to provide a best fit. Once the ethane emission is fit, we use the substantial spectral residual to provide further constraints on emission from the $H_2$ S(2) line. We note that ethane emission contributes only 15-20% of the radiance observed at 12.3 μm (814 $cm^{-1}$). This is not true of the $CH_4$ emission around the S(4) line at 1246 $cm^{-1}$, which is much more substantial, and this line was not used to constrain temperatures, although fits to this line are illustrated in **Figures 13I** and **13J**. Similar to the S(1) line, the S(3) line at 1035 $cm^{-1}$ is surrounded only by the $H_2$ CIA continuum.

>Figure 13

Initial fitting of the S(2) $H_2$ quadrupole line simultaneously with the S(1) quadrupole line proved to be impossible without a large increase of the *para*-$H_2$ ratio, which does not make any physical sense, particularly given the limited variance from the equilibrium ratio that was found by Fouchet *et al*. (2003) from matching the $H_2$ S(0) and S(1) quadrupole lines. Instead, we fit these lines by varying (i) the overall amplitude and (ii) the vertical location of the rapid temperature rise near and above the 1-μbar level that we originally patterned after the Herbert et al. (1987) UV stellar occultation results, as shown in **Figures 13A and 13B,** respectively. We did not use the analysis of these UV data by Stevens et al. (1993) because (i) their temperature profile is too warm in the stratosphere to be consistent with the IRS spectra and (ii) it is warmer than temperatures in the 10-30 μbar pressure range derived from stellar occultations (Young et al. 2001). The substantial contribution of upper-atmospheric temperatures to the upwelling radiance, peaking around $10^{-6}$ to $10^{-7}$ bars, is illustrated in **Fig. 10B**. The remaining panels show the effect of the amplitude changes and vertical location changes on the S(1), S(2), S(3) and S(4) quadrupole lines. These two perturbations together are required to match the $H_2$ S(1), S(2), S(3) and S(4) quadrupole lines simultaneously. Fortunately, these perturbations to the upper-stratospheric temperatures produce only small perturbations of the solutions in the lower stratosphere and troposphere of Uranus, and we were able to converge on a global solution for a wide range of atmospheric pressures quite rapidly. Although this obviates the possibility of solving for a stratospheric *para*-$H_2$ ratio, our assumption of a local equilibrium value is consistent with the results of Fouchet et al. (2003). Their results are worth re-examining in order to evaluate the influence of the high thermospheric temperatures on their results, although it will be smaller for the S(0) $H_2$ quadrupole line, which they considered with the S(1) line, than it is for the S(2), S(3) and S(4) lines in our analysis.

**4.3.6. Solving for the SH resolution**. Although the nominal spectral resolving power of the SH orders is R=600, we found that optimal fits to the data required adjusting the resolution within each order. Therefore, simultaneous with varying the profiles to match the data, we also vary the spectral resolution for the S(1) line at 17.04 μm (587 cm$^{-1}$) and the S(2) lines at 12.33 μm (814 cm$^{-1}$). A best fit is found for 17.04 μm (587 cm$^{-1}$), using a Gaussian spectral filter shape with a full width a half max (FWHM) of 0.023 μm (0.79 cm$^{-1}$), equivalent to a spectral resolving power R=743. A similar solution is found for both the $H_2$ S(2) line and the surrounding $C_2H_6$ lines, with a best-fit FWHM of 0.020 μm (1.33 cm$^{-1}$), equivalent to R=736. More details regarding the variation of the spectral resolution for SH orders to optimize the match to spectroscopic features will be given in Paper 2.

**4.3.7. Best-fit global-mean temperature profile**. The best-fit global-mean temperature profile is shown in **Fig. 14A** and **B**, where it is compared with several Voyager and ISO results, as well as radiative-convective models. The agreement around the tropopause is good among several data sets. In subsequent discussions and in Paper 2, we refer to our derived profile as the "nominal" profile. Alternative profiles are discussed in Section 5.1 below.

>Figure 14

## 5. Discussion

**5.1 Alternative temperature profiles.** **Figure 15** illustrates several profiles based on different starting assumptions that were perturbed to fit the data, similar to our adopted standard model. If we adopt an initial profile based on the Voyager-2 radio-occultation profile (Lindal et al. 1987, Sromovsky et al. 2011), we derive a best-fit model with a shallower temperature rise in the lower stratosphere that has a substantially poorer fit to the data, illustrating that our omission of vertical temperature oscillations in the nominal model does not degrade the fit. We also try fitting the data with a model assuming a very different bulk composition using a helium volume-mixing ratio consistent with the extreme Model G of Sromovsky et al. (2011). Except for the adiabatic extrapolation to the deep atmosphere, the retrieved temperature profile is indistinguishable from our nominal profile shown in **Fig. 15**; it is essentially the identical profile but shifted to pressures that are ~3% lower.

>Figure 15

An alternative approach that fits the spectrum just as well selects an optimum pressure level above which the atmosphere is isothermal, similar to the approach used by Fletcher et al. (2010) to model Neptune's spectral emission. This alternative best-fitting temperature profile is also shown in **Fig. 15**. Note that the isothermal portion of the atmosphere is restricted to the upper stratosphere, and that a much hotter thermosphere is required to match the $H_2$ quadrupole lines. The fact that this alternative profile can reproduce the observations to the same quality as our nominal profile indicates the existence of degenerate solutions, despite the significant constraints provided by the IRS spectra. However, **Figure 15** illustrates that our alternative profile is inconsistent with analysis of from the Voyager UVS occultation data by Herbert et al. (1987). As discussed in Paper 2, the alternative profile also requires the relative humidity to be 100% at the tropopause, which is inconsistent with Karkoschka and Tomasko (2009). Thus, we maintain the warmer stratospheric profile as our nominal model.

**5.2 Uncertainties of the derived temperatures.** The nominal 6-7% uncertainty in the absolute calibration translates into less than 0.5 K of uncertainty in the retrieved temperature. Thus, the temperature uncertainty at any given level will be dominated by systematic uncertainties and degeneracies within the family of possible solutions. The biggest known systematic uncertainty lies in the relative abundances of He and $H_2$. If we assume the 0.033 uncertainty in the He mole fraction from Conrath et al. (1987), the resulting uncertainty in retrieved temperature is limited to 1.5 K or less. One possible way to quantify the temperature uncertainties arising from the degeneracy of possible solutions could be to compare the level-by-level differences between the nominal and alternative profiles (**Fig. 15**) because both represent equally valid solutions that match the IRS data using extremes of acceptable lapse rates in the upper stratosphere. This could be considered an unrealistically conservative approach because the alternative profile is not consistent with additional observations, as discussed in Section 5.1. That said, in the well-constrained troposphere and up to pressures as low as 10 mbar, differences between the two are less than 0.5 K, increasing to 2.3 K up to the 1-mbar level, and then to 7 K at the 0.1-mbar level. At lower pressures, the alternative profile departs significantly from Voyager UVS results and this approach becomes unrealistic. Future work will need to assess and incorporate all sources of information on temperature structure to determine the appropriate uncertainties associated with the global average temperature profile for Uranus.

**5.3 Consistency with adiabatic profiles.** When we compare the derived temperature structure with adiabatic profiles, several things are noteworthy. One is that the profile is inconsistent with a $H_2$ state in which fully equilibrating para-$H_2$ and ortho-$H_2$ and their associated latent heat effects are considered at each level. The profile is most consistent with an adiabat in which para- and ortho-$H_2$ are in local equilibrium but without latent heat effects, the "frozen equilibrium" case originally proposed by Trafton (1967), together with a $CH_4$ relative humidity between 0 and 100% (**Fig. 16**), consistent with an abundance profile that will be discussed in Paper 2. The temperature profile is consistent with the radiative-convective equilibrium models of Appleby (1986) (**Fig. 14B**) in the upper troposphere.

>Figure 16

**5.4 Comparison with Voyager-2 radio-occultation results**. The derived global-mean temperatures are warmer by several degrees than the radio-occultation profile that corresponds to 2.3% $CH_4$ in the deep atmosphere derived by Lindal et al. (1987) shown in **Fig. 17A**. We cannot explain this as the result of the temperatures sampled by the radio-occultation profiles being colder than the planet as a whole. In fact, the opposite is true: Voyager-2 IRIS plots of upper-tropospheric temperatures in Uranus (Conrath et al. 1998) show that temperatures around 5°S latitude (where the radio-occultation experiment sampled) are actually warmer than those in the rest of the planet, not colder. Lindal et al. actually produced various temperature profiles that were warmer or cooler in the region below the $CH_4$ condensation level by varying its assumed abundance there and thereby changing the mean-molecular weight of the atmosphere to which the derived temperature is scaled. Their uppermost possible $CH_4$ mole fraction in the deep atmosphere is 4.4%; this profile is shown as model "F" alongside their profile "D" in **Figure 17A**.

>Figure 17

We see that the "F" profile (dotted line) matches our standard model profile at pressures between roughly 1 to 2 bars but not at pressures between 1 bar and 200 mbar; **Figures 17C** and **17E** show that the corresponding $H_2$ CIA spectrum only matches the observed $H_2$ continuum between 9.0 and 10 μm. We note that the 6-7% absolute uncertainty in radiance would correspond to brightness-temperature uncertainties in **Figs. 17C** and **17E** that correspond only to ±0.5 K or less. So increasing the assumed $CH_4$ mole fraction in the deep atmosphere does not fully resolve the differences between our results and those of Lindal et al. (1987). These differences are not resolved by appealing to contributions from distant wings of hydrocarbon emissions from $C_2H_6$ or $C_2H_2$ near 12 μm and 14 μm, respectively, because those emission features are dominated by lines emerging from the low-pressure stratosphere and are dominated by Doppler-broadened lines. We have tested that changes of pressure-broadened line widths by factors of several have negligible effect on the spectral models from these constituents.

Sromovsky et al. (2011) also addressed the issue of the mean-molecular weight of Uranus and the $CH_4$ mole fraction at depth by examining various models that fit the constraints of the vertical refractivity profile derived by Lindal et al. (1987). Some of these are illustrated in **Figure 17B**, which shows several of the temperature profiles given by Sromovsky et al. in their Fig. 6A. The various values for the $He/H_2$ ratio in these profiles (implying less He than 15%) were needed to compensate for increasing values of the $CH_4$ mole fraction in the deep atmosphere in order to match the radio-occultation refractivity profile. A $CH_4$ mole fraction of 4.44% in the deep atmosphere is the highest value that can possibly fit the refractivity profile. Sromovsky et al. found that models with the deep $CH_4$ mole fraction between 3.5 and 4.5% were most compatible with constraints placed by visible and near-infrared spectroscopic studies. **Figures 17D** and **17F** show, however, that none of these models provides a good match to the disk-averaged Spitzer
data. Although one could argue that the mole fraction of $CH_4$ could be high at the specific latitude sensed by the radio occultation, the corresponding requirement that the $He/H_2$ ratio be somewhere between 0.12 and 0.132 (Sromovsky et al. 2011), one that must be applied globally, is inconsistent with the Spitzer IRS data and radio-occultation results. Model G, with its 4.88% $CH_4$ mole fraction in the deep atmosphere can reconcile the spectra and the radio-occultation refractivity profile, but – similar to Model F of Lindal et al. (1987) – only for the 1-2 bar pressure region that is sampled between 9.0 and 10.0 μm.

The spectral match at 9.5-10.0 μm implies that reconciliation of the broader spectrum is possible by changing the mean-molecular weight throughout the atmosphere rather than just below the $CH_4$ condensation level. **Figure 18A** shows the result of making a first-order perturbation of the Lindal et al. (1987) profiles D and E results (labeled D' and E' in **Fig. 18A**) by increasing the mean-molecular weight of the atmosphere over that of the assumed 15/85 $He/H_2$ ratio, plus level-by-level contributions from $CH_4$, by 4.0% and 3.5%, respectively. We note that this is only a first-order perturbation because it ignores smaller changes in the pressure levels that are necessary in re-interpreting the refractivity profile, as did Sromovsky et al. (2011). Nonetheless, such an exercise is useful in exploring possible solutions. The results are interesting: these models now fit the 9 – 12 μm continuum (the "heavy" version of model E in particular), although they result in higher brightness temperatures than measured for the $H_2$ CIA continuum in the 16 – 20 μm region (**Fig. 18E**). Let us assume, for the moment, that those higher temperatures at 400 mbar and lower pressures (**Fig. 18A**) are a result of the radio-occultation experiment sampling of one of the warmest regions on the planet near 5°S latitude. An increase of the mean-molecular weight of 3.5% would correspond to a higher $He/H_2$ ratio, i.e. 19/81, an increase in the He mole fraction of 0.04. This value is 1.2 times the 1-σ uncertainty of 0.033 cited by Conrath et al. (1987) and therefore a value of 0.19 for the He mole fraction is not numerically implausible. That said, we do not expect the $He/H_2$ ratio greater than solar, because enrichment mechanisms, such as the one suggested by Guillot and Hueso (2006), are effective only for atoms much heavier than helium.

>Fig. 18

On the contrary, one might argue on cosmogonic grounds that the He/H$_2$ ratio of 15/85 is close to that of the solar nebula (Asplund et al. 2009). It is more plausible that He would be depleted with respect to hydrogen because of phase separation of helium and subsequent formation of heavy helium-rich droplets that sediment to deeper levels, as has been suggested originally by Stevenson and Salpeter (1977). An alternative approach to increasing the mean molecular weight would be to increase the abundance of a heavier molecule that is spectroscopically unlikely to be detected, e.g. N$_2$ or Ar. In order to retain the He/H$_2$ ratio of 15/85, but increase the mean-molecular weight by 3.5%, one would need N$_2$ to be present at a mole fraction of 6.9 x 10$^{-3}$, which is more than an order of magnitude greater than predicted by the chemical models of Fegley and Prinn (1986) ($< 1.3$ x 10$^{-4}$), or Ar to be present at a mole fraction of 2.1 x 10$^{-3}$, which is many orders of magnitude greater than any plausible formation scenario (Conrath et al. 1987). Such a large abundance might also lead to detectable abundances of HCN through galactic cosmic ray impacts on N$_2$ and further chemistry (Lellouch et al. 1994), which is not observed. Thus, we can offer no completely satisfactory solution to this problem at this time, but note that any further constraints on the bulk composition of Uranus, particularly those that provide tighter constraints on the He/H$_2$ ratio, will enable some clarification of this discrepancy, which deserves further exploration in the future.

We can perturb the radio-occultation profiles within the limits of their uncertainties, estimated as 2K around the 2-bar level and 1K around the 0.1-bar level by Lindal et al. (1987), independent of compositional uncertainties. **Figure 18D** illustrates the profile of Model F1 of Sromovsky et al. (2011) and the profile with a 1-K temperature increase ("F1+1K"), which is well within the limits of these uncertainties. The resulting H$_2$ CIA spectra are shown in **Figures 18D** and **18F**. The F1+1K profile is still colder than the measured spectrum and its uncertainties at the short-wavelength end, but is warmer than the measured spectral continuum from 570 to 610 cm$^{-1}$. There are no net improvements to the fit for larger offset values. A vertically variable offset profile, such as one with an offset of +2 K at the 2-bar level and -1 K at the 0.1-bar level, does not work; these offsets are averaged out in the spectrum with little net change.

We note that determinations of the mean-molecular weight and He/H$_2$ ratio from comparisons of Voyager radio-occultation and thermal-infrared observations have proven to be problematic in the past. The helium abundance derived from that technique needed to be revised upward at Jupiter to match Galileo probe results (von Zahn et al. 1998, Niemann et al. 1998) and at Saturn to match the infrared-only analysis (Conrath and Gautier 2000). The presence of such systematic errors in the Voyager measurements, which remain unexplained, is at least qualitatively consistent with the need for a higher mean-molecular weight to reconcile the Voyager radio-occultation results and the Spitzer IRS spectra.

Another way to resolve this disagreement is to postulate a source of stratospheric opacity, such as a particulate layer, that provides the additional continuum opacity. We test the effectiveness of stratospheric particulates to reconcile the IRS and radio-occultation results by placing additional opacity in a layer at several levels in the stratosphere and varying the optical thickness until the spectrum corresponding to Sromovsky et al.'s Model F1 temperature profile with an additional 1 K of warming matches the observed 11-μm radiance. For a layer at 100 μbar (122.7K, assuming our derived temperature structure for the upper stratosphere), the required optical thickness is $1.5 \times 10^{-5}$, with an approximate uncertainty of 50%. For a layer at 31.6 μbar (138.6K), it is $5.0 \times 10^{-6}$, and for a layer at 10 μbar (155.2K), it is $2.0 \times 10^{-6}$, both with ~50% uncertainties. We do not consider hazes at lower pressures, because these pressures are already located at higher altitudes than both the methane homopause with its associated complex hydrocarbon production region and the predicted water condensation region from any external oxygen source, so there are no known sources of hazes at such high altitudes. Moreover, there is no evidence for stratospheric hazes at pressures lower than ~0.1 mbar in the high-phase angle Voyager-2 imaging data (Pollack et al. 1987, Rages et al. 1991). These required optical depths cam be compared with those observed at visible wavelengths from the Voyager-2 imaging analysis, where Pollack et al. (1987) and Rages et al. 1991) found that at the total optical thickness of the haze column that extends up to the formation altitude is on the order of $1 \times 10^{-4}$. Their simulation of the particle properties in this region included values for the particle radii that ranged from 0.068 down to 0.011 μm. If these particles are predominantly scattering, their optical thickness at longer wavelengths are lower by an inverse fourth-power of the wavelength; if they are predominantly absorbing, they are lower by the simple inverse of the wavelength (Hansen and Travis 1974). Assuming the largest particles in this range, extrapolation to the 10-μm region implies a total optical thickness of only $2 \times 10^{-13}$ for Rayleigh scattering but $7 \times 10^{-7}$ for Rayleigh absorption, both too small to match the smallest optical thickness required to reconcile the IRS data and Model F1+ 1 K.

On the other hand, if the particles have a low imaginary index of refraction in the visible and high imaginary index near 10 μm, it is possible for them to have a significant infrared optical thickness while having been essentially invisible to the Voyager imaging system. **Figures 18D** and **18F** illustrate the $H_2$ CIA + haze spectrum for the assumed additional haze at 31.6 μbar, compared with the observations and our best-fit model. Assuming that the optical thickness of the haze varies spectrally as the inverse power of the wavelength, consistent with Rayleigh absorption, the resulting model (long-dashed line) is too bright in the 15-20 μm region and closer to the observed spectrum but lower than the uncertainty of the measurements in the 9-10 μm region. Although inconsistent with the assumption that the particle are highly absorbing, for completeness we also tested the assumption that the haze varies spectrally with the inverse fourth-power of the wavelength, consistent with Rayleigh scattering. This model spectrum (dotted line) does better at 9-10 μm but is still substantially higher than the observational uncertainties at 15-20 μm. Although not shown for clarity in **Fig. 18D**, the model spectrum for a 100-μbar haze layer does considerably worse, and a 10-μbar haze layer does not improve the fit over the 31.6-μbar haze-layer model.

However, we cannot absolutely rule out the possibility of significant contributions by stratospheric hazes to the spectrum. We consider here only a single-layer, spectrally uniform test, and many other particle distributions and particulate spectra could exist that would match the spectral continuum. It would be worth investigating this idea further with realistic aerosol microphysical models and extinction calculations from appropriate materials to predict more accurately the expected particle size distributions, particle densities, and extinction properties of hazes produced, say, from recondensed meteoric-ablation debris. The observations of stratospheric $H_2O$ and $CO_2$ could help constrain the total available amount of condensable material, given estimates of the rock-ice fraction in Kuiper-belt dust or other source material. One means to determine whether warm hazes are contributing to the emission is to measure the distribution of the upwelling radiance as a function of planetary radius. **Figure 19** illustrates the differences between 11-µm continuum radiances from our standard model and Model F of Lindal et al. (1987), which are nearly identical, and the test using Model F1 of Sromovsky et al. (2011) + 1K and the stratospheric particulate layer whose spectrum is shown in **Figs. 18D** and **18F**, which is measurably less limb darkened. The faintness of the spectrum at this wavelength makes such observations difficult or impossible from ground-based observatories, but it would be accessible to space-based cryogenic platforms.

>Figure 19

**5.5. Comparison with Voyager infrared data**. **Figure 14B** includes a comparison of the temperatures derived from the Voyager IRIS experiment (Conrath et al. 1998) for the warmest and coolest regions on the planet. Our nominal profile lies in between them.
**Figure 20** shows a more direct comparison between the data from which these temperature profiles were derived and the spectra generated by our standard model for the same mean emission angles for each region. The spectrum from our profile is correspondingly colder than the spectrum for the warmest region (**Fig. 20A**), and warmer than the spectrum for the coldest region (**Fig. 20B**). The somewhat greater similarity of our model spectrum to the warmest spectrum than the coolest spectrum can be attributed to the geometry of the planet at the equinox, in which the warm low-latitude regions contribute significantly to the global mean. Because temperatures just below the tropopause up through lower pressures could be subject to some seasonal variability (Friedson and Ingersoll 1987, Conrath et al. 1990), a closer examination of the seasonal dependence of temperatures at various latitudes is warranted as a part of more detailed comparisons of these two data sets. **Fig. 20** also illustrates that the "high-methane" temperature profiles corresponding to Model F of Lindal et al. (1987) and to Model F1 of Sromovsky et al. (2011) and the same model with a 1-K uniform warming all produce spectra that are too cold for the warm near-equatorial IRIS data, which is the latitude region that the radio-occultation experiment sampled. This demonstrates that the difficulty with these models is not confined to the Spitzer IRS data alone, at least without major changes in the meridional variation of temperature structures between the Voyager-encounter epoch and late 2007.

>Figure 20

Figure 14A compares our results with analyses of the Voyager UVS occultation experiments (Herbert et al. 1987, Stevens et al. 1993). As noted earlier, we favor the Herbert et al. profile over those of Stevens et al. The profiles developed by Stevens et al. at pressures greater than ~1 µbar from the UV occultations at $H_2$ Rayleigh scattering wavelengths are too warm to be consistent with our data. The Stevens et al. profiles are similar to the warmest temperature profile in **Fig. 14a**, which illustrates the various stratospheric lapse rates tested in our temperature retrievals. The Stevens et al. profiles result in over three times the residuals to the fit to the $H_2$ continuum compared with using the Herbert et al. profile. Furthermore, the Stevens et al. profile is also too warm to be consistent with temperatures derived from stellar occultations that sense temperatures in the 10-30 µbar pressure range (Young et al. 2001). Although our thermospheric solution is sensitive to the assumed temperatures at stratospheric altitudes that are not well constrained by the Spitzer data (see Section 5.1 and **Fig. 15**), we note that our nominal profile is consistent with the temperatures in the Herbert et al. profile in the stratosphere but is colder in the thermosphere. Although tempting to attribute this to differences between local occultation measurements vs. the global mean to which our data are sensitive, the UVS measurements were taken at 63.7ºS, 4ºS and 69.7ºN, covering a wide latitude range. On the other hand, this temperature difference is also consistent with a net cooling of thermospheric temperatures between the 1986 epoch of Voyager measurements to our 2007 Spitzer measurements. Such a drop in temperatures would be a drop in ionospheric temperatures detected by the 1992-2011 monitoring of $H_3^+$ emission described by Melin et al. (2013), although they point out that the ostensibly colder temperatures could also be explained by changes in the orientation of the magnetic field with respect to the solar wind that allows electrons to be accelerated to higher energies and thus deposited deeper into colder regions of the thermosphere. Nonetheless, a drop in temperatures would also be consistent with a drop in temperatures detected by stellar occultations between 1983 and 1998 (Young et al. 2001). Because of the extremely long radiative time scales characterizing this part of the atmosphere (Conrath et al. 1990), such a cooling implies some form of dynamical energy transport at work.

**5.6. Implications for longer wavelengths and standard spectral models**.
Extrapolating the spectrum of the nominal model to the far-infrared and submillimeter wavelengths should provide a useful basis for a physical model for this region. This is important, in part, because Uranus is used as a part of the absolute calibration system of the Herschel Space Telescope (Pilbratt et al. 2010). **Figure 21** shows the spectrum derived from our best-fit model using only the CIA continuum, together with existing observations in this spectral region. A spectrum based on the Voyager radio occultation temperature profile corresponding to Model D of Lindal et al. (1987), although too cold for the middle infrared (e.g. **Fig. 15B**), is nonetheless roughly consistent with the observations in this region (see the lower dotted-dashed line in **Fig. 21** and Fig. 4 of Griffin and Orton 1993). Our nominal model matches the middle infrared and is within the uncertainties of far-infrared and submillimeter filtered radiometric points at wavelengths shorter than 0.8 mm (**Fig. 21**), although the Lindal et al. (1987) Model F and Sromovsky et al. (2011) Model F1 do slightly better. However, our nominal model is too warm at wavelengths longer than 0.8 mm (wavenumbers shorter than 12.5 cm$^{-1}$). Griffin and Orton (1993) noted that another opacity source was needed to match the longest point on this graph at 3.3 mm, and they fit this point by adding opacity from $NH_3$, which is expected in the deeper atmosphere, giving it a mole fraction of $1 \times 10^{-6}$ and a simple falloff above the condensation level governed by the saturation vapor pressure. However, using our warmer nominal temperature profile, the opacity of $NH_3$ required to match the 3.3-mm point does not add enough opacity to lower the spectrum sufficiently to match data between 0.8 and 3.3 mm (3 and 12.5 cm$^{-1}$) (dashed line in **Fig. 21**). To a lesser extent, this is also true of the Lindal et al. Model F and Sromovsky et al. Model F1.

>Figure 21

One solution is to invoke another molecular opacity source. The solid line in **Fig. 21** illustrates the spectrum where the opacity of $H_2S$ is added to those of $H_2$ and $NH_3$ in the Uranus radiative-transfer model. We use the absorption model developed by de Boer (1995), which was used to model the atmosphere of Neptune (de Boer and Steffes 1996). We adjust the $H_2S$ abundance in the deep atmosphere to match the 3.3-mm point (mole fraction equal to $1.4 \times 10^{-5}$) and a "humidity" and scale height above its saturation level to match the 0.8-3.3 mm data (a value of 1.6 times the atmospheric scale height). This model spectrum illustrates that $H_2S$ must be considered seriously as a candidate opacity source, just as de Boer and Steffes (1996) considered it for Neptune and de Pater et al. (1991) considered it for the microwave spectrum of both Uranus and Neptune. Although less likely on the grounds of relative solar abundances, $PH_3$ should also be investigated as a candidate opacity source. The spectrum that includes $H_2S$ in **Fig. 21** shows that moderate-resolution spectroscopy (R ~ 30 or better) in the 0.8-4 mm (2.5 – 12.5 cm$^{-1}$) region could detect the presence of $H_2S$ with a signal-to-noise ratio that is better than that of the Serabyn and Weisstein (1996) spectrum, shown by the open circles.

**6. Summary, Conclusions and Future Work.**

We present four sets of spectra of Uranus between 7.4 and 19.6 μm, equally spaced in longitude. In addition to covering the $H_2$ collision-induced opacity "continuum" and quadrupole lines, spectroscopic signatures are detected of emission from vibration-rotation bands associated with $CH_4$, $CH_3D$, $C_2H_2$, $C_2H_6$, $CH_3C_2H$, $C_4H_2$, $CO_2$ and possibly $CH_3$. Although parts of the spectrum dominated by collision-induced absorption by $H_2$ do not differ above an expected measurement uncertainty of 3%, we note a significantly higher variability of hydrocarbon emission over a rotation of the planet. Future work can determine whether the longitudinal variability must be explained by correlated variations of hydrocarbon abundances or by variations of stratospheric temperature.

We analyze the average of these spectra to derive temperatures using CIA opacity and quadrupole-line emission of well-mixed $H_2$. The derived lapse rate between 1 and 2 bars of pressure is consistent with adiabatic conditions, and that assumption is also true for pressures as low as 500 mbar under the assumptions of a $CH_4$ relative humidity that is subsaturated. Our choice for a nominal temperature profile fits the IRS $H_2$ CIA and quadrupole constraints simultaneously and is consistent with the analysis of Voyager UVS data by Herbert et al. (1987) and with the analysis of Infrared Space Observatory Short-Wavelength Spectrometer measurements of S(0) and S(1) $H_2$ quadrupole lines by Fouchet et al. (2003). A best fit to the S(1), S(2), S(3) and S(4) $H_2$ quadrupole lines is obtained by perturbing the very high temperature profile from the microbar to the nanobar pressure regime derived by Herbert et al. (1987). This fit assumes equilibrium mixtures of the abundances of para- vs. ortho-$H_2$, consistent with the results of Fouchet et al. (2003). Additional constraints on this ratio could be provided by IRS Long-High module observations of the S(0) $H_2$ quadrupole, but significant corrections in the high-resolution order containing this feature must be made with the continuum fit to a model.

Our derived temperatures are higher than the Voyager-2 radio-occultation profiles (Lindal et al. 1987) and than the best-fitting revised profile of Sromovsky et al. (2011) that assume low (~2%) methane missing ratios in the deep atmosphere. Models with higher values, such as those with a 4% volume mixing ratio, do much better but still do not match significant portions of the $H_2$ CIA continuum. We have no definitive reason at this time why this is the case, although we are sampling the planetary average and the radio-occultation profiles measured only low latitudes. We note that a partial solution to the difference could be attributed to a heavier mean molecular weight than implied by the Voyager-derived results (Conrath et al. 1987) with a He mole fraction of 0.19. Although this He abundance is only 1.2 σ higher than the Conrath et al. (1987) results, such a high value is problematic because it is larger than the solar ratio, and there are no plausible mechanisms published to date for enhancing the He/$H_2$ ratio over solar values. Alternatively, the temperature difference could arise from a component such as $N_2$ that has not yet been identified spectroscopically, although theoretical considerations do not favor this interpretation. Further insight into this inconsistency would be provided by independent constraints on the bulk composition; these might revise upward the Voyager results for the mean-molecular weight, as was the case for Jupiter and Saturn in comparison with independent data (e.g., Conrath and Gautier 2000). The models of Sromovsky et al. (2011) that require a lower He/$H_2$ ratio than derived from Voyager radio-occultation and infrared data are inconsistent with our observations, Their models could be made consistent if an additional absorber, such as a particulate layer, were placed in the stratosphere. Although we cannot absolutely rule out such a particulate layer, our tests require it to have particular spectral properties: a low imaginary index of refraction in the visible that rendering it undetectable by the Voyager imaging system (Pollack et al. 1987, Rages et al. 1991), a high imaginary index of refraction in the 11-µm region that makes it an effective emitter, with a sharp reduction of its emissivity toward longer wavelengths. This possibility of stratospheric emission of any kind can be tested by imaging or spatially resolved spectroscopy near 11 µm, because a stratospheric emission results in distinctly less limb darkening than is otherwise the case.

Another source of systematic error could lie in the model for the $H_2$ CIA absorption. Orton et al. (2007) cite a 5% uncertainty in the coefficients for wavenumbers up through 1200 cm$^{-1}$; a 5% decrease in the CIA absorption would make the brightness-temperature spectrum in the 800 – 1100 cm-1 range warmer by 0.2 – 0.6 K. This would make the Sromovsky et al. (2011) Model F1+1K close to the measurements in the 1000 – 1100 cm$^{-1}$ region, but they would remain far from the measurements, including uncertainties, at 870 – 1000 cm$^{-1}$. Changes to the brightness-temperature spectrum at shorter wavenumbers are less than 0.2 K and would not mitigate any of the differences between the models and the measured spectra. The CIA absorption would need to be "corrected" by ~30-35% to match the observed radiances in the 870 – 1000 cm$^{-1}$ region for the F1+1K model. Although it is a difficult measurement, we repeat the recommendation of Orton et al. (2007) for laboratory measurements in this region to verify the absorption model in this spectral range.

Further work should be done to constrain the solutions for temperatures. Radiative-convective models would provide an additional constraint on families of models whose stratospheric temperature profiles are also tied to $CH_4$ mole-fraction profiles because of methane's central role in radiative heating. These models would update those of Friedson and Ingersoll (1997) and Conrath et al. (1990), ideally merging models of energy balance in the stratosphere with those for the thermosphere, extending the work of Herbert et al. (1987) and Stevens et al. (1993). Future model constraints will be derived from the Herschel Space Telescope (Hartogh et al. 2009). Comparisons with these will be addressed in future work, e.g. the profile of Uranus derived by Feuchtgruber et al. (2013) based on constraints from HD line measurements by Herschel and ISO will be addressed in future work. These models and analysis of Herschel and ground-based observations will provide a valuable context to the sensitive, spatially resolved measurements expected from the James Webb Space Telescope (Gardner et al. 2006). JWST measurements could provide the center-to-limb test for the presence of a stratospheric emitter in the very faint 11-µm region.

Comparisons of our nominal model with submillimeter through microwave observations imply that that an additional source of opacity is required besides the $H_2$ CIA continuum and $NH_3$ inversion lines; candidates for such opacity include $H_2S$ or $PH_3$. With such an opacity in place and the good fit to observations at shorter wavelengths, the model would have a practical use: establishing Uranus as a key component in a system of calibration standards, particularly for the far-infrared through millimeter spectrum where other reliable sources provide inadequate flux, such as for Herschel Space Telescope instruments (e.g. SPIRE, Swinyard et al. 2010). Thus, it is an important objective of our immediate future work to pull all these lines of inquiry together with the goal of a unified disk-averaged flux model, based on physical properties, whose goal is an accuracy that is better than 5%. Such a model should be reinforced by further observations in the far infrared through millimeter spectral region and expanded to include spatial and temporal variability. High-precision measurements in the 0.3 to 0.8 mm wavelength range could also provide an independent discriminator between temperature models around the 70-80 K (~0.8 – 1 bar) level (**Fig. 21**).

On-line Supplemental Materials for this paper include all the reduced IRS data for all modes that we used in this paper. Supplemental Materials for Paper 2 include the nominal temperature profile, along with the vertical profiles of minor and trace constituents developed in that paper.

**Acknowledgements**.

We thank NASA's Spitzer Space Telescope program for initial support of the data acquisition, reduction and its initial analysis, and we thank Tom Soifer for Director's Discretionary Time on Spitzer (program #467). This work is based on observations made with the Spitzer Space Telescope, which is operated by the Jet Propulsion Laboratory, California Institute of Technology under a contract with NASA. Support for this work from the Spitzer program was provided by NASA through an award issued by JPL/Caltech. Another portion of our support was provided to JPL/Caltech, from NASA's Planetary Atmospheres program. J. Moses acknowledges support from NASA grants

NNX13AH81G, as well as older grants from the NASA Planetary Atmospheres program. L. Fletcher acknowledges the Oak Ridge Association of Universities for its support during his tenure at the Jet Propulsion Laboratory in the NASA Postdoctoral Program (NPP), together with the Glasstone and Royal Society Research Fellowships during his current tenure at the University of Oxford. F. J. Martin-Torres acknowledges support from the Spanish Economy and Competitivity Ministry (AYA2011-25720 and AYA2012-38707). During his contribution to this work, M. Line was supported by NASA's Undergraduate Student Research Program (USRP).

This research made use of Tiny Tim/Spitzer, developed by John Krist for the Spitzer Science Center. The Center is managed by the California Institute of Technology under a contract with NASA.

The radiative-transfer calculations were primarily performed on JPL supercomputer facilities, which were provided by funding from the JPL Office of the Chief Information Officer.

We thank Linda Brown, Helmut Feuchtgruber, Tristan Guillot, Mark Hofstadter, Kathy Rages, Larry Sromovsky and Larry Trafton for helpful and illuminating conversations, J. Schaefer for help in implementing dimer contributions into the $H_2$ collision-induced opacity calculations, Emmanuel Lellouch and an anonymous reviewer for helpful comments and suggestions, and Larry Sromovsky for digital versions of various temperature profiles presented by Sromovsky et al. (2011).

# Tables

**Table 1.** Module characteristics summary table. No LL1 data (19.5-38.0 μm) were acquired because of expected detector saturation in these long wavelengths.

| Module | Wavelength Range, λ (μm) | Spectral Resolution, R=λ/Δλ | Slit Width (arsec) | Point-Spread Halfwidth (arcsec) |
|---|---|---|---|---|
| SL1 | 7.46-14.05 | 60-120 = 8.2667λ | 3.7 | 1.9-3.5 |
| SL2 | 5.13-7.52 | 86-144 = 16.533λ | 3.6 | 1.1-1.9 |
| LL2 | 13.98-21.43 | 85-126 = 5.9048λ | 10.5 | 3.6-4.3 |
| SH | 9.95-19.30 | 600±60* | 4.7 | 2.4-4.7 |
| LH | 19.27-35.97 | 600±60* | 11.1 | 4.7-9.3 |

*The error on spectral resolution for the high-resolution modules (given by the Spitzer Observers Manual) is 1σ.

**Table 2.** Observation summary table. Longitude numbers refer to the sequence shown in Table 3. Uranus was sampled twice at each longitude; the standard observation sequence included "nodding" the object, i.e. observing it in two positions on the slit.

| Module | Integration Time (s) | Number of Integrations | Total Time on Source (s) | Longitudes |
|---|---|---|---|---|
| SL1 | 14 | 25 | 734 | 1-4 |
| SL2 | 14 | 25 | 734 | 1-4 |
| LL2 | 6 | 15 | 189 | 1-4 |
| SH | 120 | 15 | 3657 | 1-4 |
| SH | 120 | 5 | 1219 | N/A-sky |
| LH | 6 | 4 | 50 | 1-4 |
| LH | 6 | 2 | 25 | N/A-sky |

**Table 3**. Observing time associated with each module. Because Spitzer must be repointed to observe with each different spectrograph module, the observing times vary with longitude. For this table, we assume a 17.24-hour rotational period, as adopted by the International Astronomical Union's Working Group on Cartographic Coordinates and Rotational Elements. (Archinal et al. 2011).

| Module | Longitude | UT Date | UT Time | Time Relative to Start (hrs) | Longitude (°West, System III) |
|--------|-----------|---------|---------|------------------------------|-------------------------------|
| SL | 1 | 2007 Dec.16 | 18:00 | 0.0 | 208.3 |
| SL | 2 | 2007 Dec. 17 | 00:24 | 6.40 | 342.0 |
| SL | 3 | 2007 Dec. 17 | 04:57 | 10.95 | 77.0 |
| SL | 4 | 2007 Dec. 17 | 13:33 | 19.55 | 256.5 |
| LL | 1 | 2007 Dec. 16 | 19:38 | 1.63 | 242.4 |
| LL | 2 | 2007 Dec. 17 | 02:04 | 8.07 | 16.8 |
| LL | 3 | 2007 Dec. 17 | 08:44 | 14.73 | 156.0 |
| LL | 4 | 2007 Dec. 17 | 13:38 | 19.63 | 258.3 |
| SH | 1 | 2007 Dec. 16 | 19:01 | 1.02 | 229.6 |
| SH | 2 | 2007 Dec. 17 | 01:27 | 7.45 | 3.9 |
| SH | 3 | 2007 Dec. 17 | 08:00 | 14.00 | 140.6 |
| SH | 4 | 2007 Dec. 17 | 14:36 | 20.60 | 278.5 |
| LH | 1 | 2007 Dec. 16 | 19:43 | 1.72 | 244.2 |
| LH | 2 | 2007 Dec. 17 | 02:08 | 8.13 | 18.2 |
| LH | 3 | 2007 Dec. 17 | 08:49 | 14.82 | 157.7 |
| LH | 4 | 2007 Dec. 17 | 14:32 | 20.53 | 277.0 |

# Figures

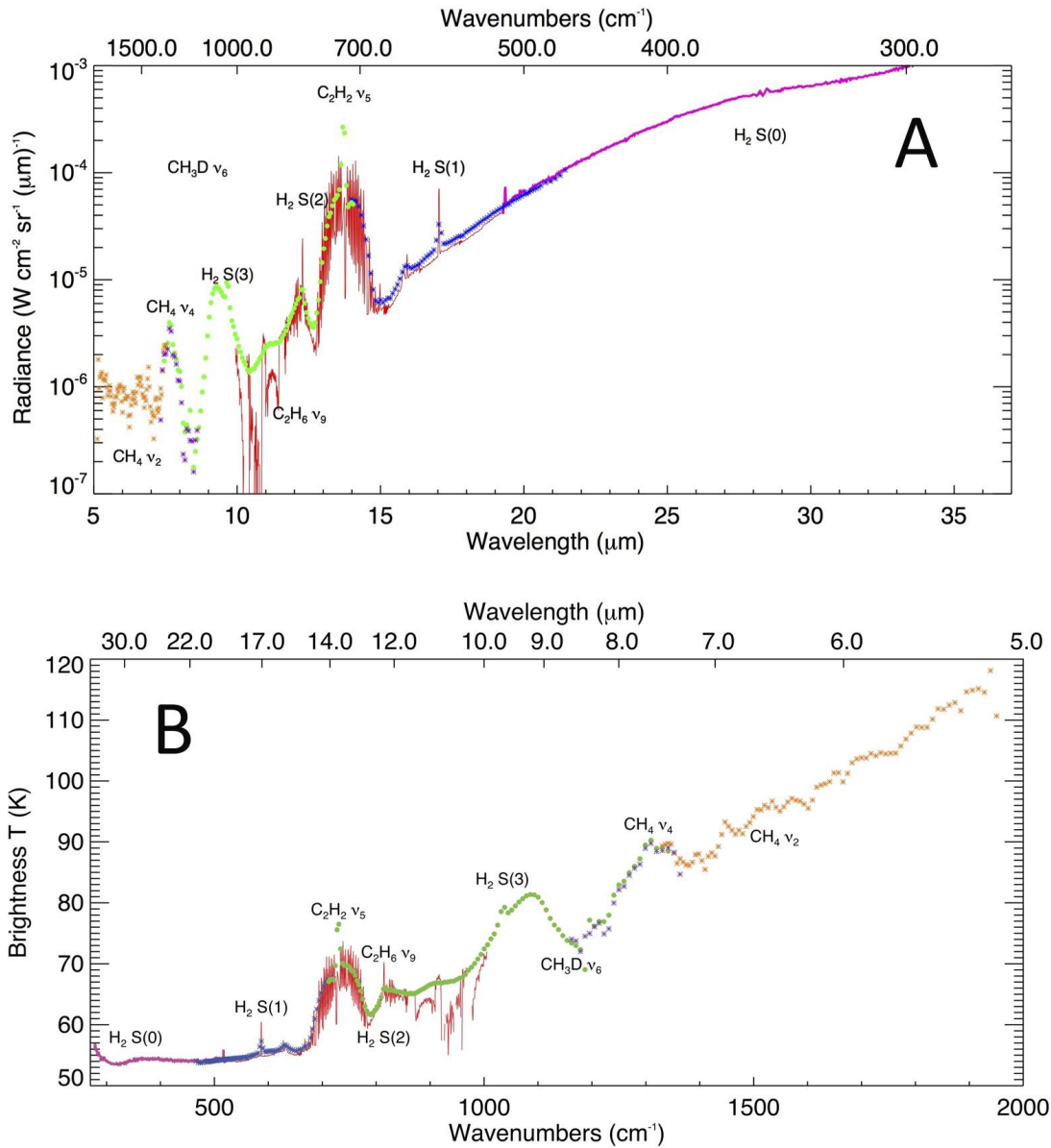

**Figure 1**. The spectrum of Uranus, based on the mean of samples at all nod positions and longitudes, as observed by all modules the Spitzer IRS on 2007 December 16-17. Different modules are coded in different colors and shapes: SL2 - orange asterisks, SL "bonus order" – purple asterisks, SL1 - green points, LL2 – blue asterisks, SH – red lines, LH – purple points. *Panel A*. Radiance *vs.* wavelength. *Panel B*. Equivalent brightness temperature for radiance vs. wavenumber, with wavelength given by the upper abscissa. Rendition in brightness temperature allows the spectral regions that sample temperatures at different atmospheric levels to be identified more readily. Some spectral features are identified for clarity in each panel.

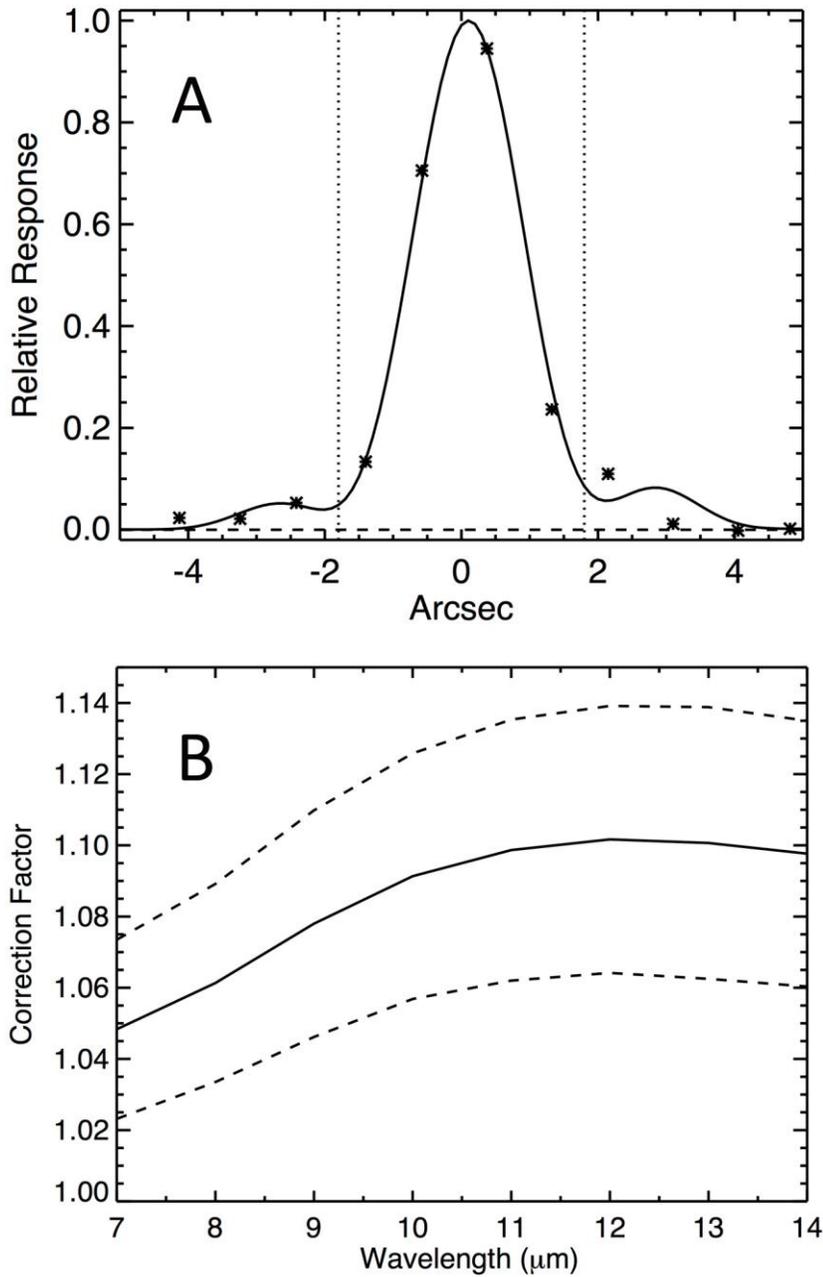

Figure 2. Corrections for overfilling the SL slit. *Panel A*. Comparison of a scan across the SL slit for Uranus at 14.0 μm (asterisks) and a model scan that was created from the convolution of a model point-spread function and a homogeneously radiating ("flat") disk of the size of Uranus (solid line). The dashed line indicates zero. Vertical dotted lines indicate the edges of the SL slit. *Panel B*. Schematic of the spectral dependence of the factor needed for SL observations to correct for overfilling the slit with the disk of Uranus. The increase with wavelength is the result of diffraction; the rolloff at the longest wavelengths is because flux from the stellar standard is also beginning to be lost from the slit. The dashed lines represent uncertainties arising from the unknown center-to-limb behavior within the disk of Uranus

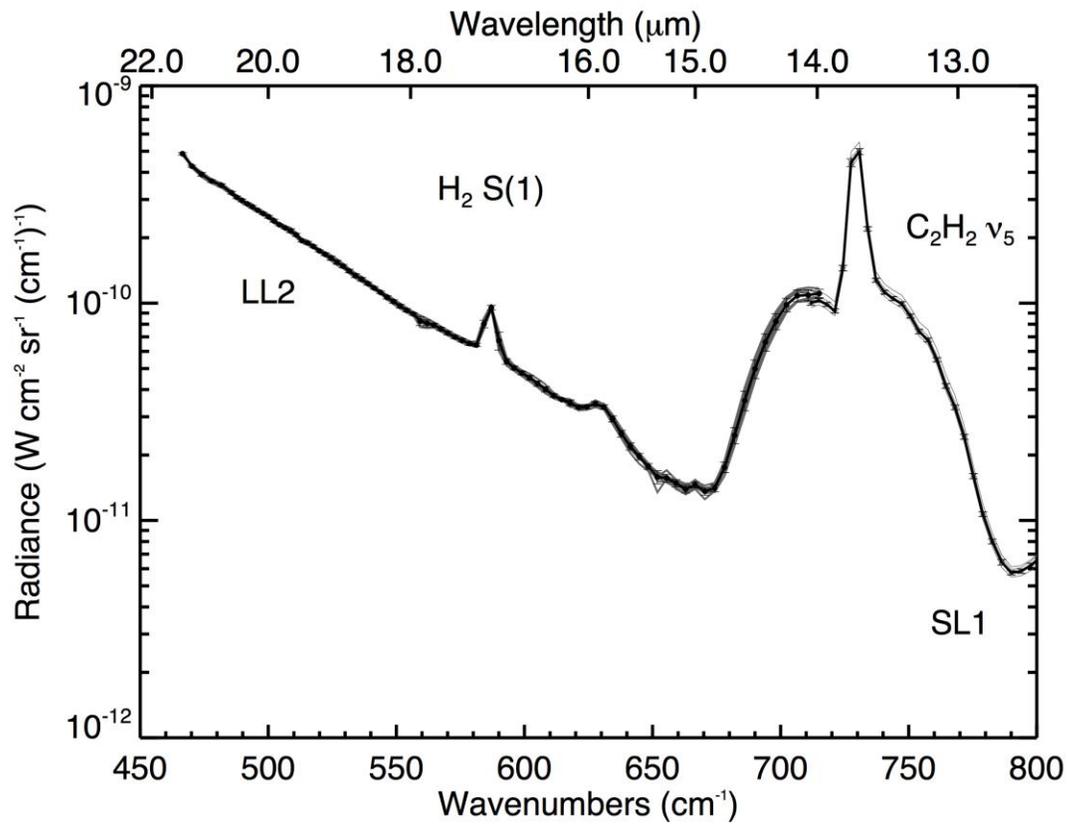

**Figure 3**. Comparison of the mean SL1 spectrum (right side: crosses and solid black line) and the mean LL2 spectrum (left side: filled circles and solid black line). A small inconsistency in the overlap region of the acetylene emission feature near 14 μm is probably the result of averaging different samples of longitudinal variations in the two different modules. The individual spectra comprising the mean are also shown by the gray lines.

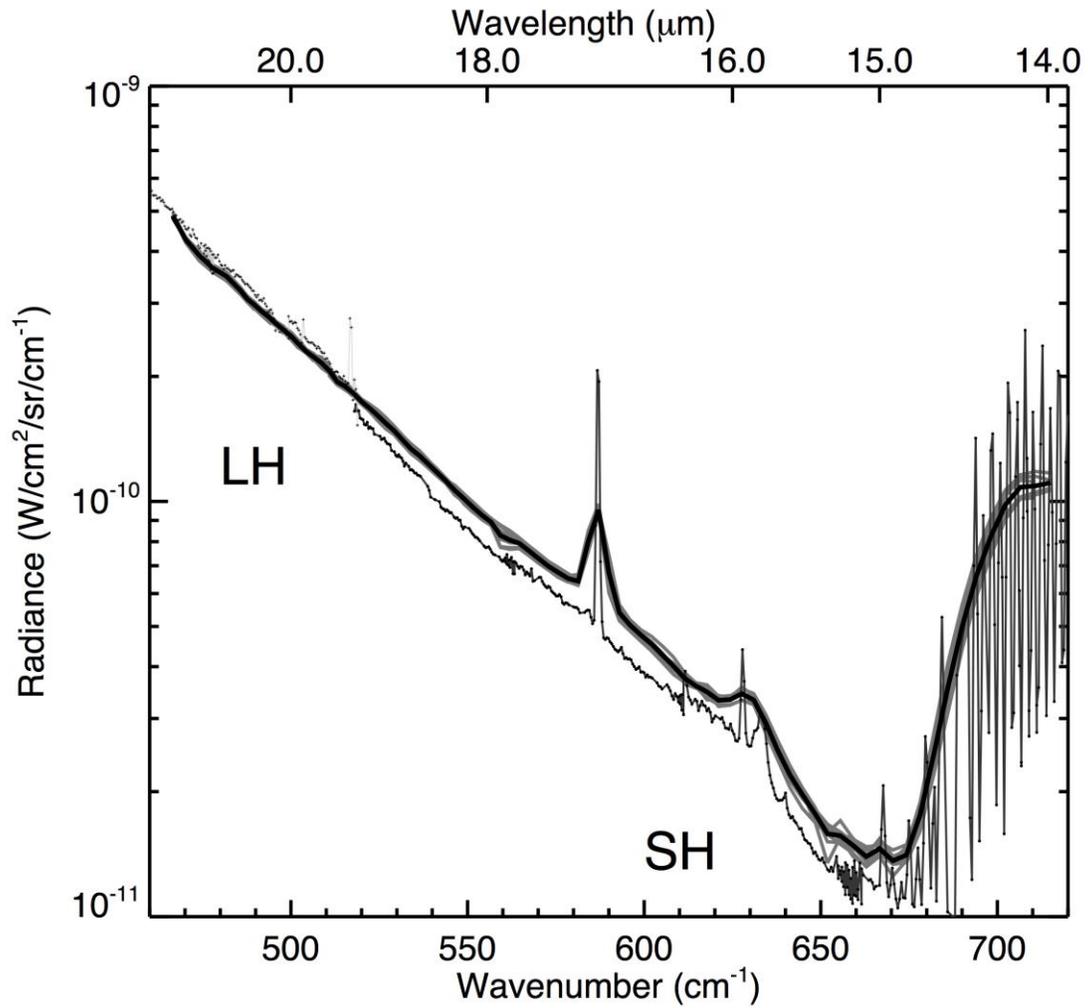

**Figure 4**. Comparison of the LL2 spectrum (black line) with SH spectra (light gray line with filled circles for wavelengths less than 19.42 µm; wavenumbers greater than 515 cm$^{-1}$) and LH spectra (light gray line with filled circles for wavelengths greater than 19.42 µm; wavenumbers less than 515 cm$^{-1}$). Individual LL2 spectra constituting the mean are also shown by gray lines around the LL2 mean.

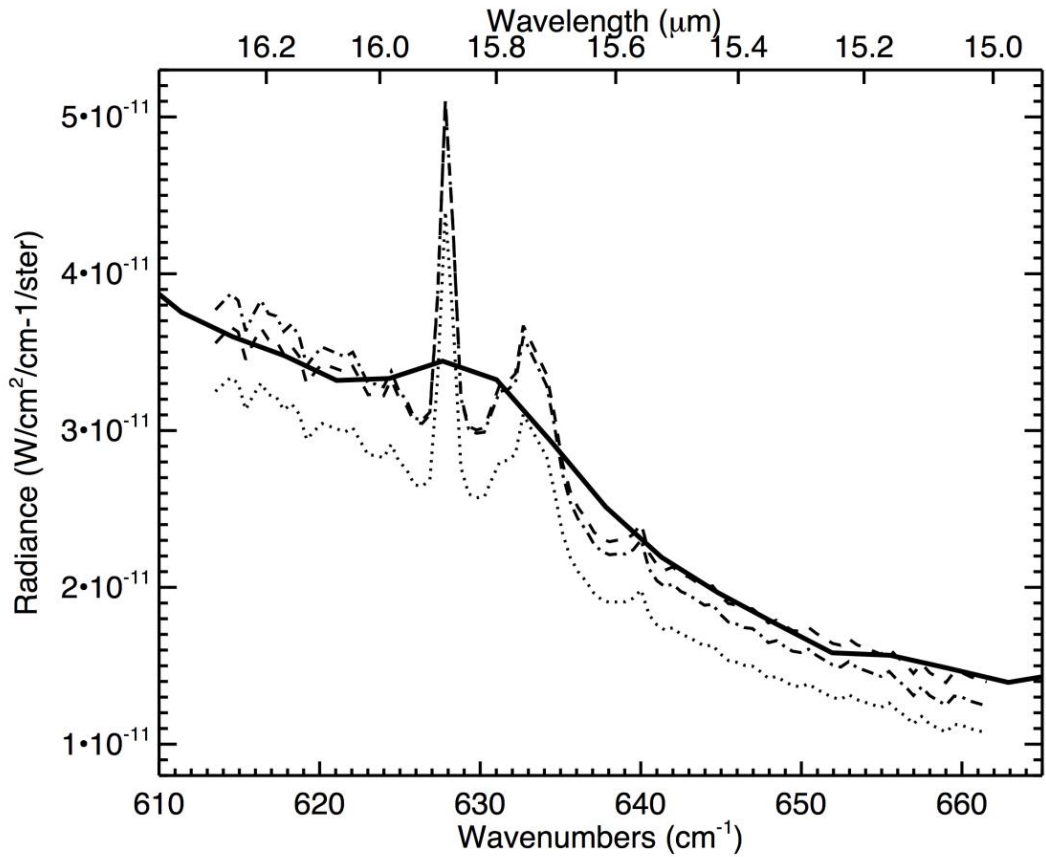

**Figure 5**. Example of scaling and "pivoting" required for SH spectra to match the more photometrically reliable low-resolution spectra. In this case, the SH order-13 spectrum (lower dotted curve) requires not only upward rescaling (dotted-dashed curve) but also the addition of a constant spectral slope, "pivoting" (thick dashed curve) to match the LL spectrum (thick solid line).

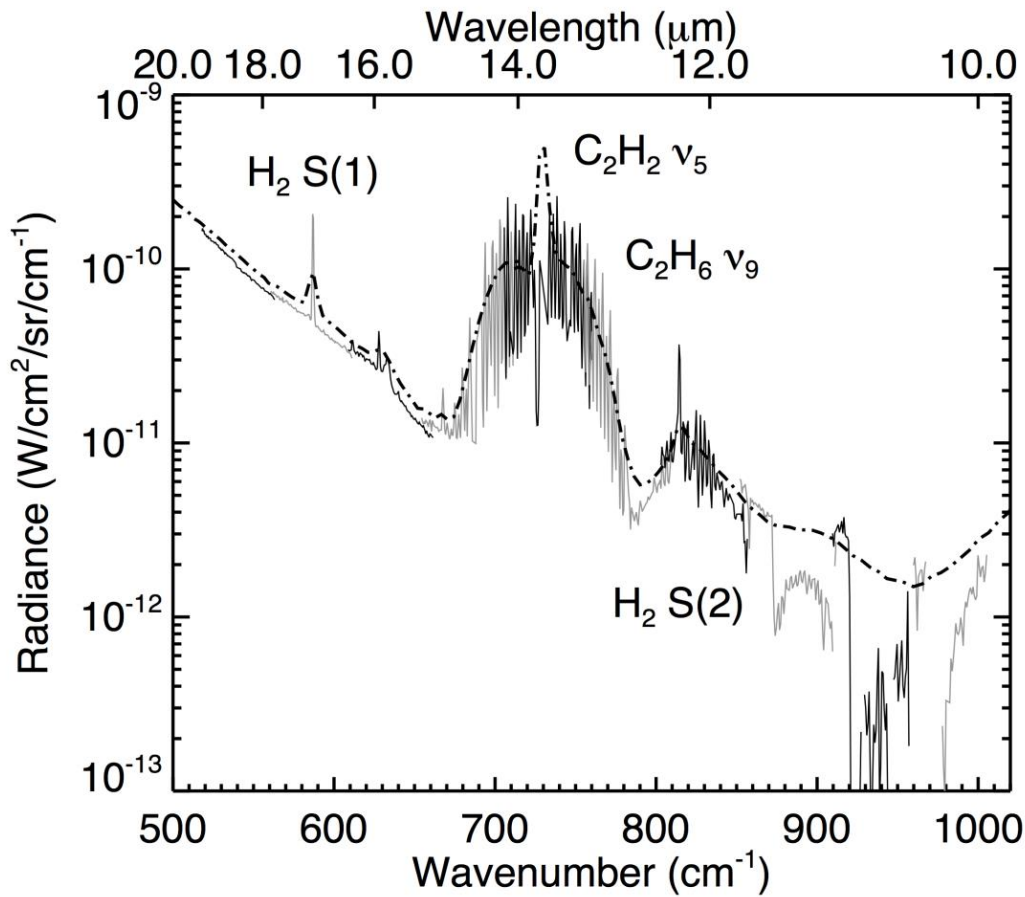

**Figure 6**. Comparison of SH, SL1 and LL2 spectra. The individual SH spectral orders, coadding the nods and longitudes for each, are denoted by alternating black and gray lines, going from order 11 on the left to order 20 on the right. Despite careful selection of 'good' data points at the end of each order, this plot demonstrates that inconsistencies still exist. All nods and longitudes are also coadded for SL1 and LL2 spectra, denoted in the dark alternating dashed-dotted line. The calibration of the low-resolution spectra is expected to be more reliable than the high-resolution spectra because of their internal consistency. The SH orders show both internal inconsistencies and offsets from adjacent orders, making the spectra extremely difficult to interpret.

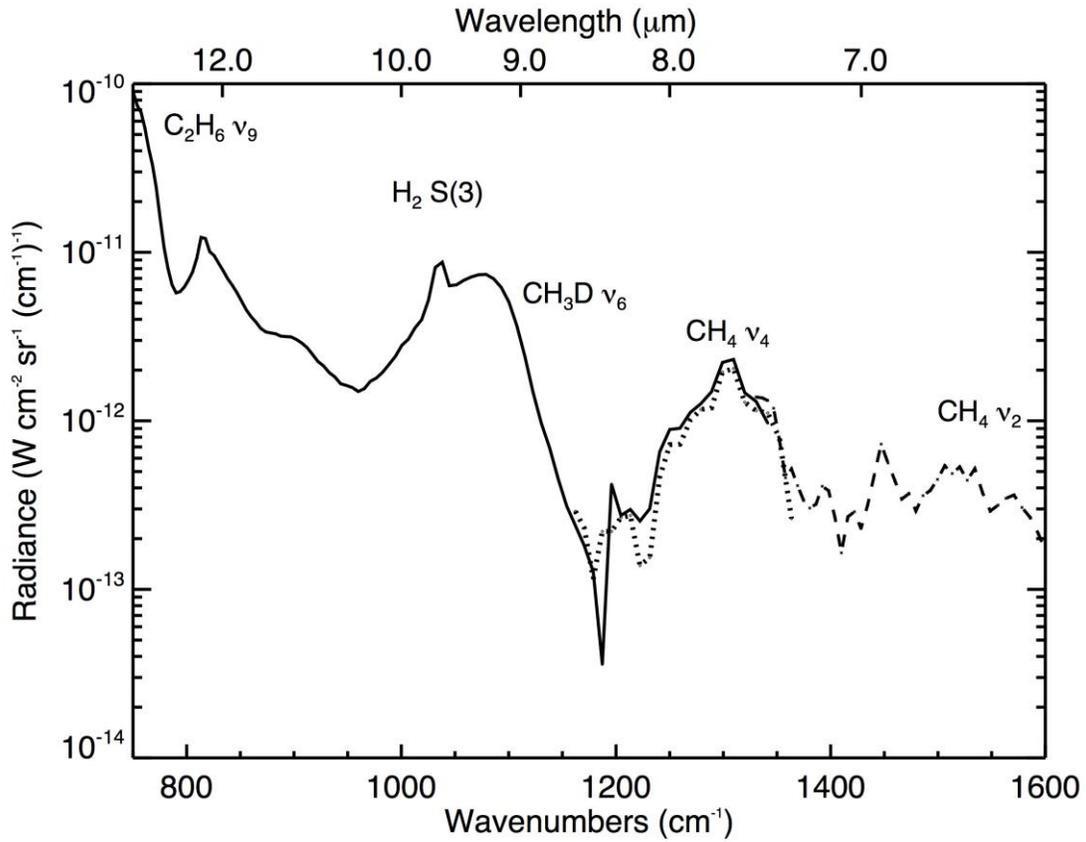

**Figure 7**. Comparison of the coadded SL1 (solid line), SL2 (dashed line), and bonus (dotted line) orders. Some differences between the modules are apparent at the peak of the $CH_4$ emission at 7.66 μm (1306 cm$^{-1}$).

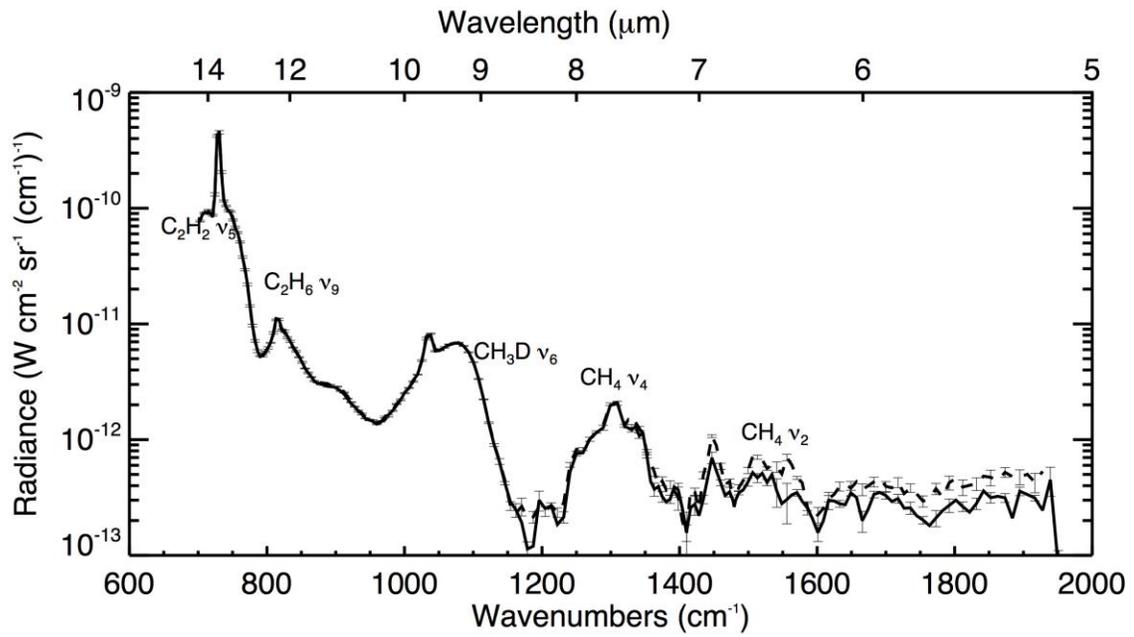

**Figure 8**. Comparison of the mean short-wavelength, low-resolution spectra from the 2007 Director-Discretionary observations discussed in this paper (solid curve) with the equivalent spectra in our 2005 Cycle-1 observations (dashed curve). Error bars sample the measurement uncertainty in each spectrum.

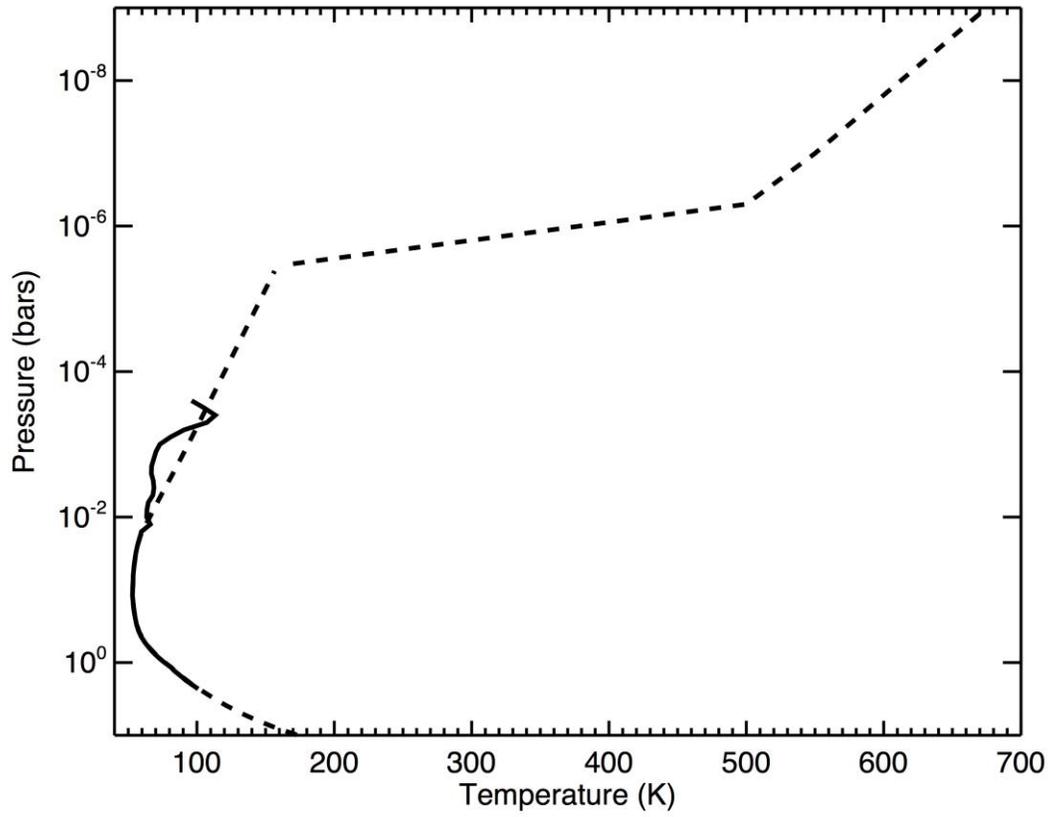

**Figure 9**. Initial temperature structure assumed. We smooth through the oscillations of the radio occultation profile (Lindal et al. 1987), both because we cannot resolve such waves by inverting IRS data, and because they may vary across the disk.

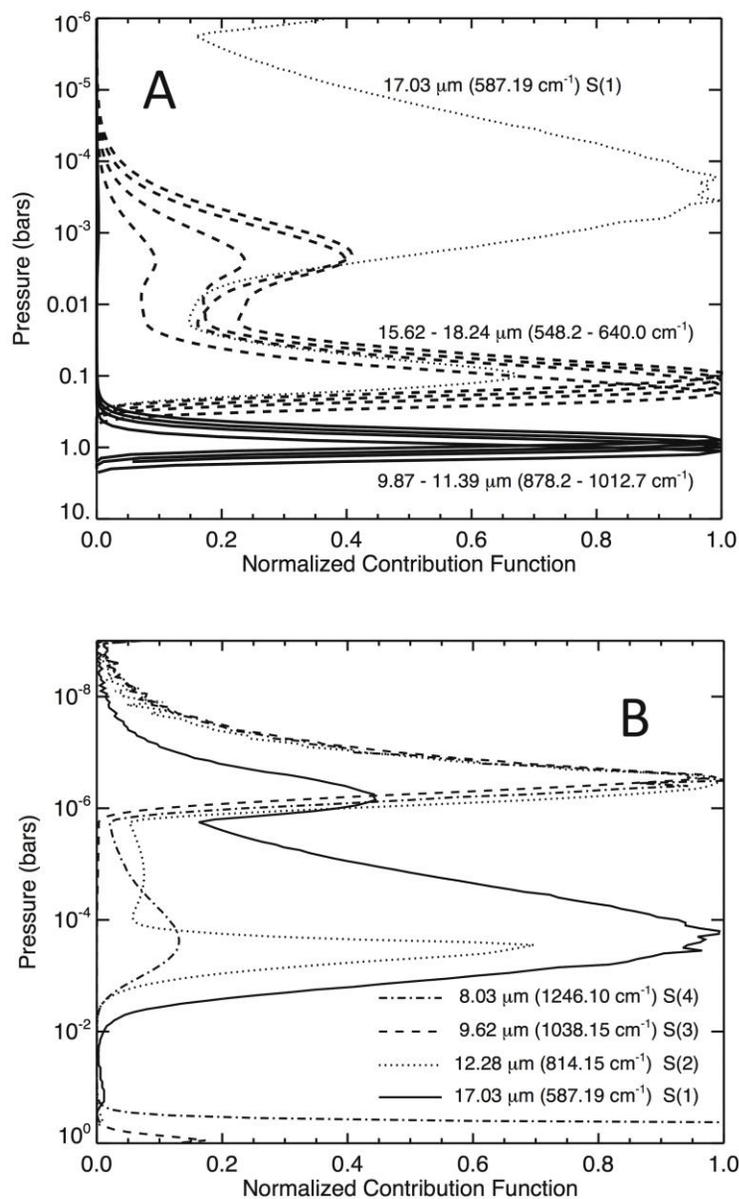

**Figure 10**. Contribution functions associated with wavelengths used to derive the temperature profile. *Panel A*. Sample display of normalized contribution functions for spectral points along the $H_2$ collision-induced continuum in the (dashed lines) 15.63-18.25 μm (548-640 cm$^{-1}$) and (solid lines) 9.88-11.36 μm (880-1012 cm$^{-1}$) regions of the spectrum, together with the peak of the $H_2$ S(1) quadrupole at 17.035 μm (587.04 cm$^{-1}$). *Panel B*. Normalized contribution functions for spectral points at the peaks of the $H_2$ quadrupole lines at the spectral resolution of the IRS data. The S(1) and S(2) characterized at high resolution and S(3) and S(4) at low resolution. Because much of the total outgoing radiance at these spectral points arises from the $H_2$ collision-induced continuum, and from the emission of $C_2H_6$ lines for the $H_2$ S(2) feature, we isolate the stratospheric component by subtracting a contribution function for the nearest spectral continuum.

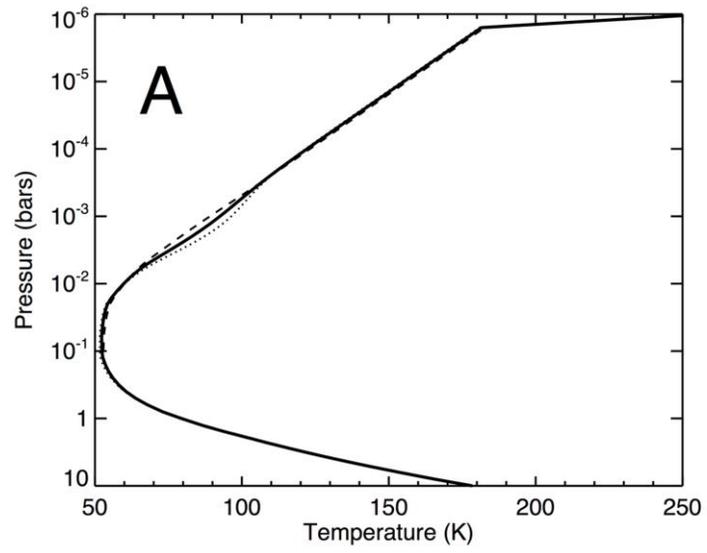
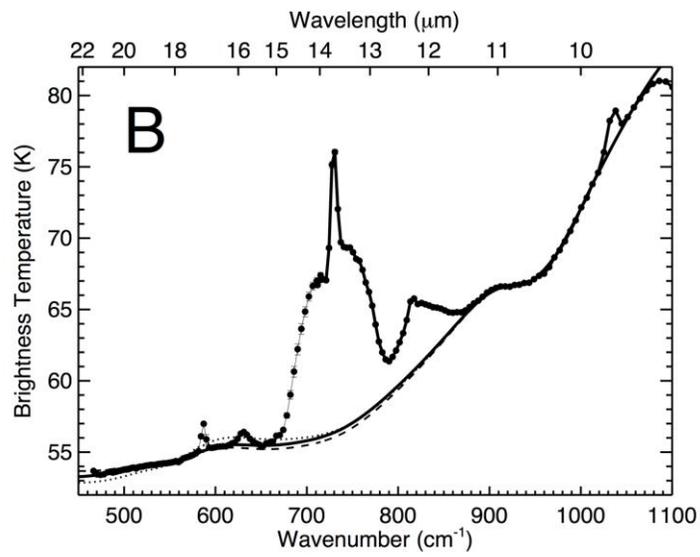
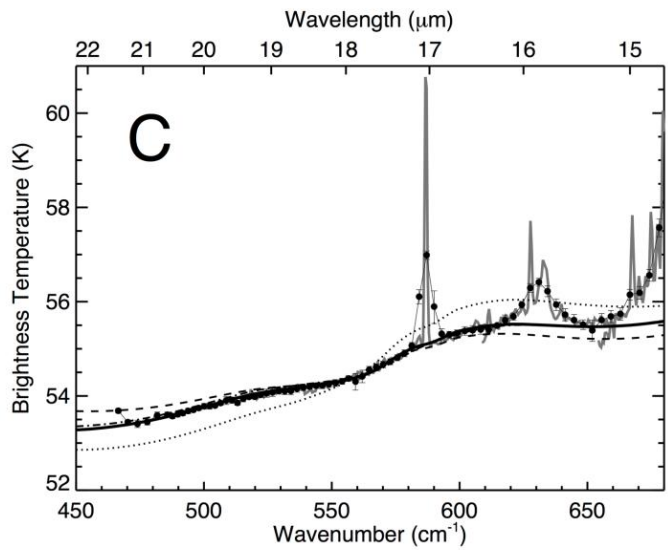

**Figure 11**. (previous page) Optimizing the fit of temperatures in the lower stratosphere. *Panel A*. Perturbations to lower stratosphere that were used to optimize the fit to the spectra shown in Panels B and C. The best-fit model is given by the solid line. *Panel B*. The thick solid line shows the best-fit model to the spectrum, from perturbations to the lower stratosphere shown in Panel A. SL1 and LL2 spectra are shown as the thin solid line with filled circles.. *Panel C*. Details of the 14.3-22.2 μm (450-700 cm$^{-1}$) spectra that are shown in Panel B. Scaled SH data are included in this panel and shown as the gray solid line with the LL spectra shown as the black solid line. The dot-dashed line in Panel C denotes the spectrum of the nominal model (solid line) if the influence of dimers is not included. Note that we do not fit the emission features to derive the temperature profile, only the H$_2$ collision-induced continuum between them. Note that in this and subsequent figures showing the low-resolution spectra, measurement error bars are displayed but are often smaller than the

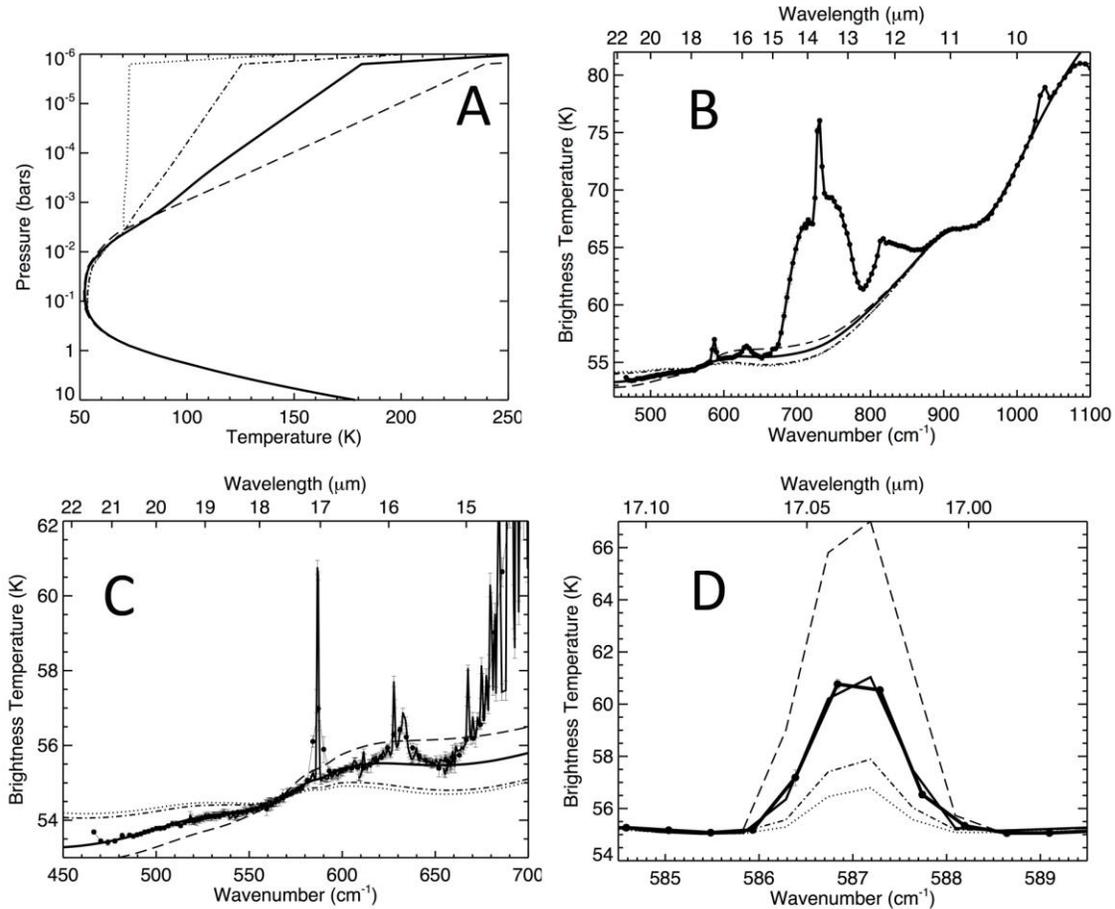

**Figure 12**. Optimizing the fit to the lapse rate of the upper stratosphere. *Panel A*. Perturbations to the lapse rate of the upper stratosphere. The best fit is shown by the model temperature profile with the solid line. *Panel B*. Sensitivity of the $H_2$ collision-induced continuum spectrum to the perturbations of these temperature profiles. *Panel C*. Details of the 14.3-22.2 μm (450-700 cm$^{-1}$) spectra that are shown in Panel B. Data are shown as in **Fig. 11**. As in **Fig. 11**, we are only fitting the $H_2$ collision-induced absorption continuum between emission features. *Panel D*. Sensitivity of the $H_2$ S(1) quadrupole emission spectrum to the same temperature perturbations. The linestyles for the different models in Panels B, C and D correspond to those used in Panel A.

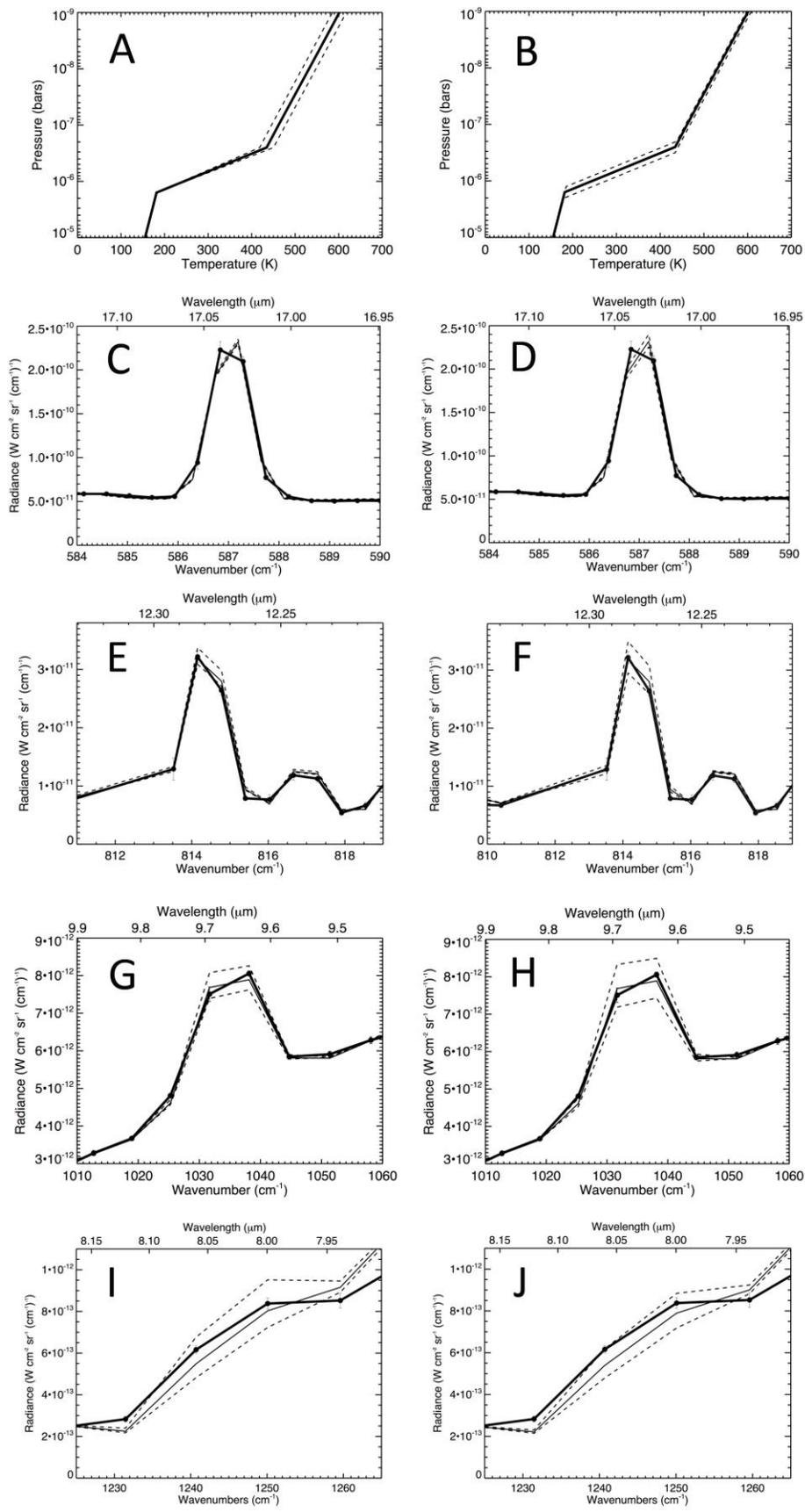

**Figure 13** (previous page). Effect of perturbations of the temperature profile of Herbert et al. (1987) on spectra of $H_2$ quadrupole lines. *Panel A* shows the amplitude perturbations of the profile. *Panel B* shows perturbations of its vertical location. The solid curve represents the best-fit solution to the $H_2$ quadrupole lines shown in the remaining panel. The dashed curves represent the ±1 standard deviation limits of the goodness of fit ($\chi^2$), determined by the values corresponding to the minimum value of $\chi^2+1$ (see, for example, Bevington and Robinson 1992). *Panels C, E, G* and *I* (left column) illustrate the sensitivity of the $H_2$ S(1), S(2), S(3) and S(4) quadrupole lines, respectively, to the temperature perturbations shown in Panel A. *Panels D, F, H* and *J* (right column) illustrate the sensitivity of the $H_2$ S(1), S(2), S(3) and S(4) quadrupole lines, respectively, to the temperature perturbations shown in Panel B.

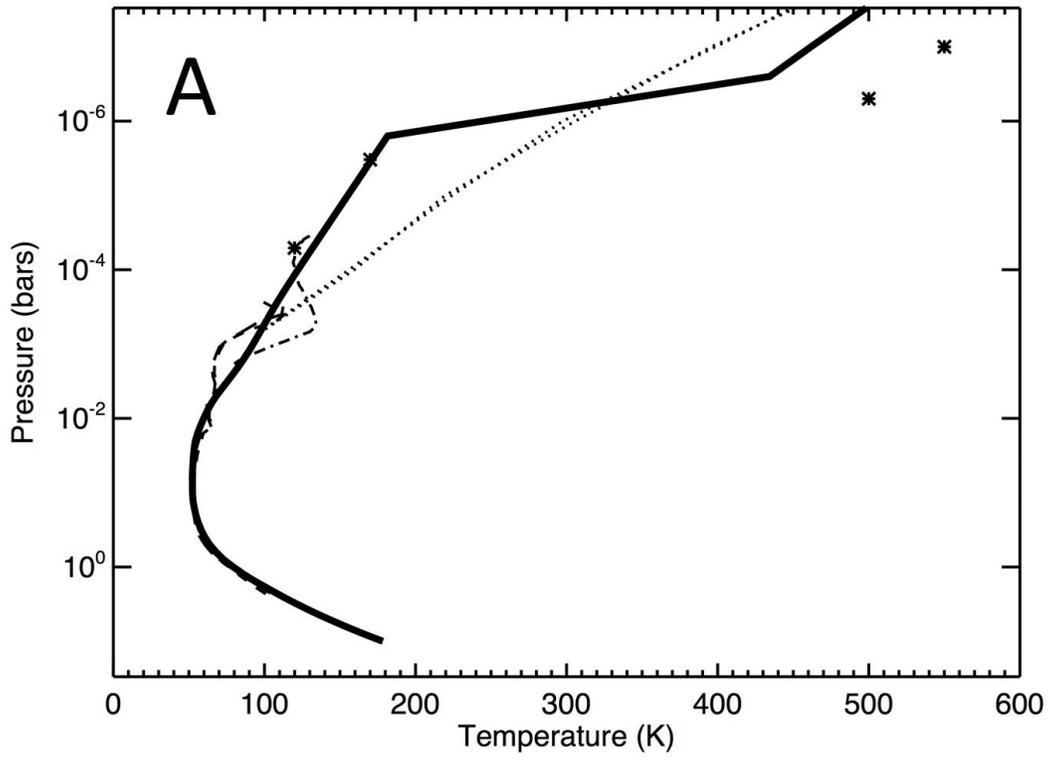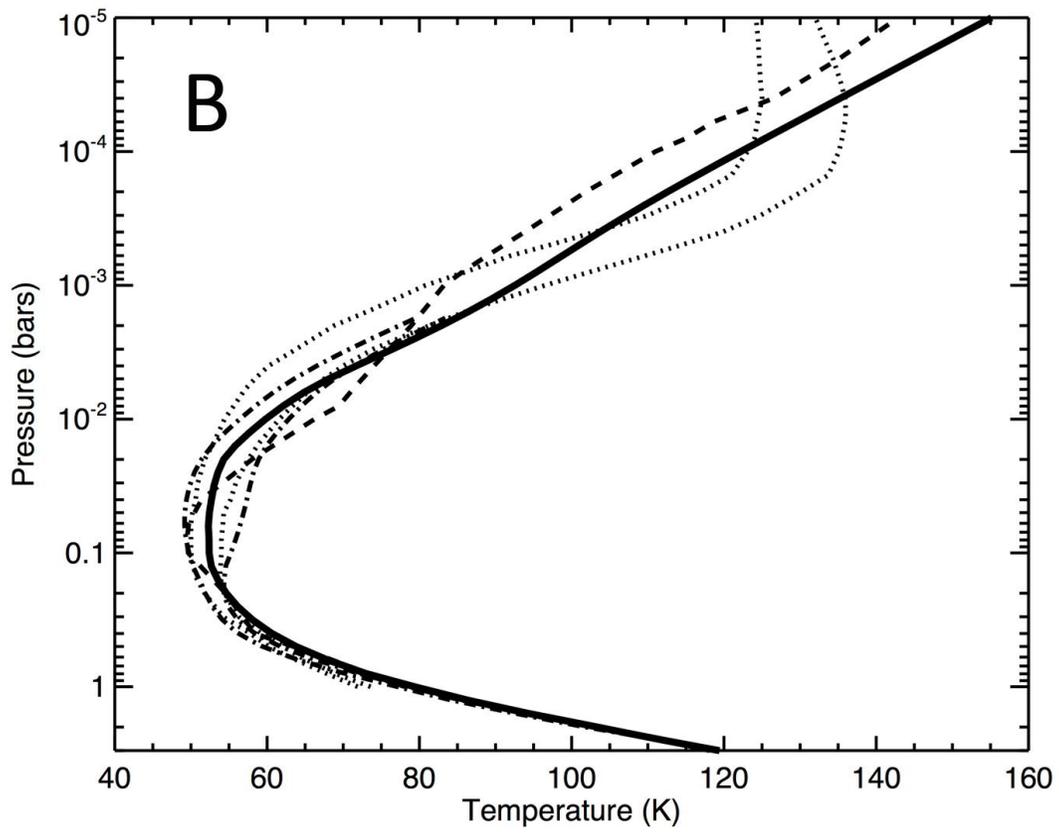
50

**Figure 14**. (previous page) Comparison of temperature profiles. The thick solid line indicates our nominal profile in both panels. *Panel A*. Asterisks represent published points along the compromise temperature profile developed by Herbert et al. (1987). The short-dashed lines represent the Voyager radio-subsystem profile (Model F) of Lindal et al. (1987) and the adjacent long-dashed lines the Voyager radio-subsystem profile (Model F1) of Sromovsky et al. (2011). The dotted-dashed line represents the results of Bishop et al. (1990) and the dotted line the results of Stevens et al. (1993). *Panel B*. This detailed plot of the higher-pressure region shows a comparison with the temperature profile derived by Fouchet et al. (2003) from ISO measurements of the $H_2$ S(0) and S(1) quadrupole transitions that is given by the dashed line. The dotted lines illustrate the range of radiative-convective equilibrium models of Appleby (1986). The dotted-dashed lines show the warmest and coldest temperature profiles derived from the Voyager IRIS experiment (Conrath et al. 1990).

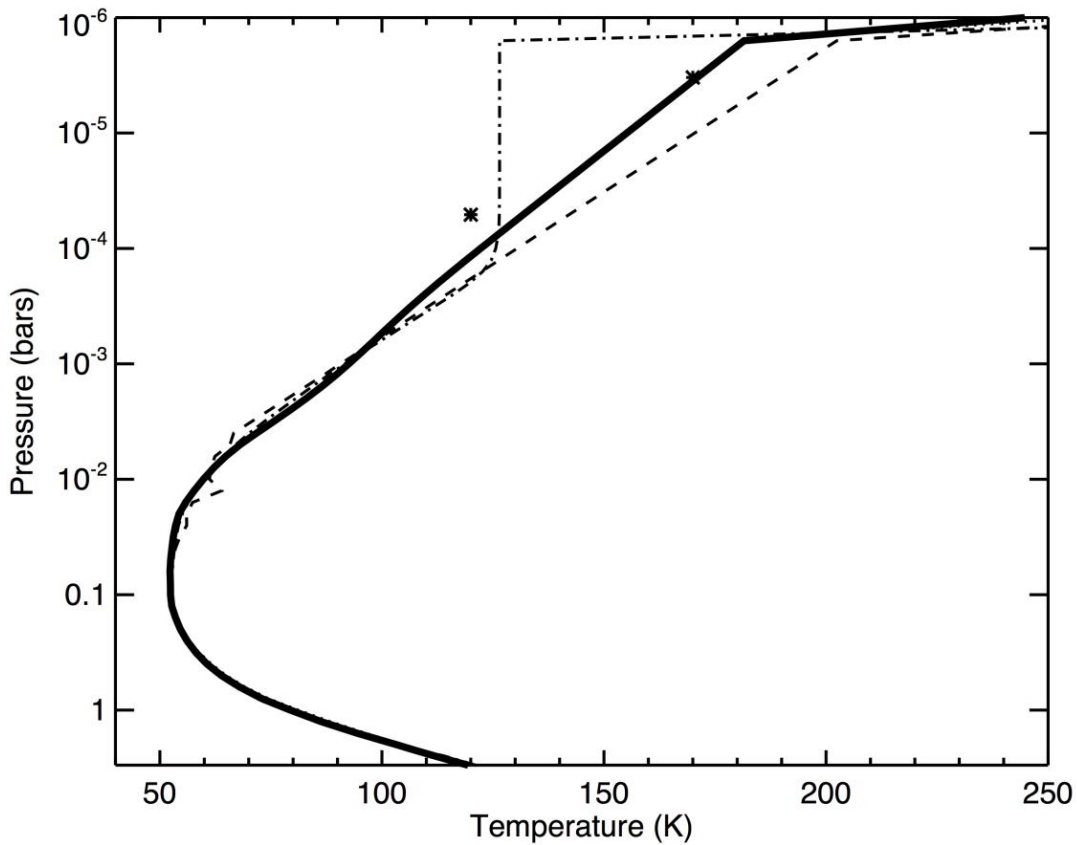

**Figure 15.** Alternative temperature profiles. The thick solid line shows the standard profile. The dashed line shows a model with similar perturbations as displayed in **Figs. 12** and **13** but beginning with the radio occultation profile given by Model D of Lindal et al. (1987). The dotted line, distinguishable from the standard model in this graph only for pressures greater than 0.2 bars, was retrieved in the same way as the standard model but assuming a much lower helium volume-mixing ratio of 0.1063, consistent with the extreme Model G of Sromovsky et al. (2011). The dashed-dotted line illustrates an alternative temperature profile whose lower stratosphere is assumed to be isothermal between 0.1 mbar and the pressure associated with the thermospheric temperature increase. Although the spectrum associated with this profile is indistinguishable from that of the best-fit standard model shown in **Figs. 11, 12** and **13**, it is much colder than the temperatures derived from the analysis of Voyager UVS occultation observations using a double heating source by Herbert et al. (1987) given by asterisks.

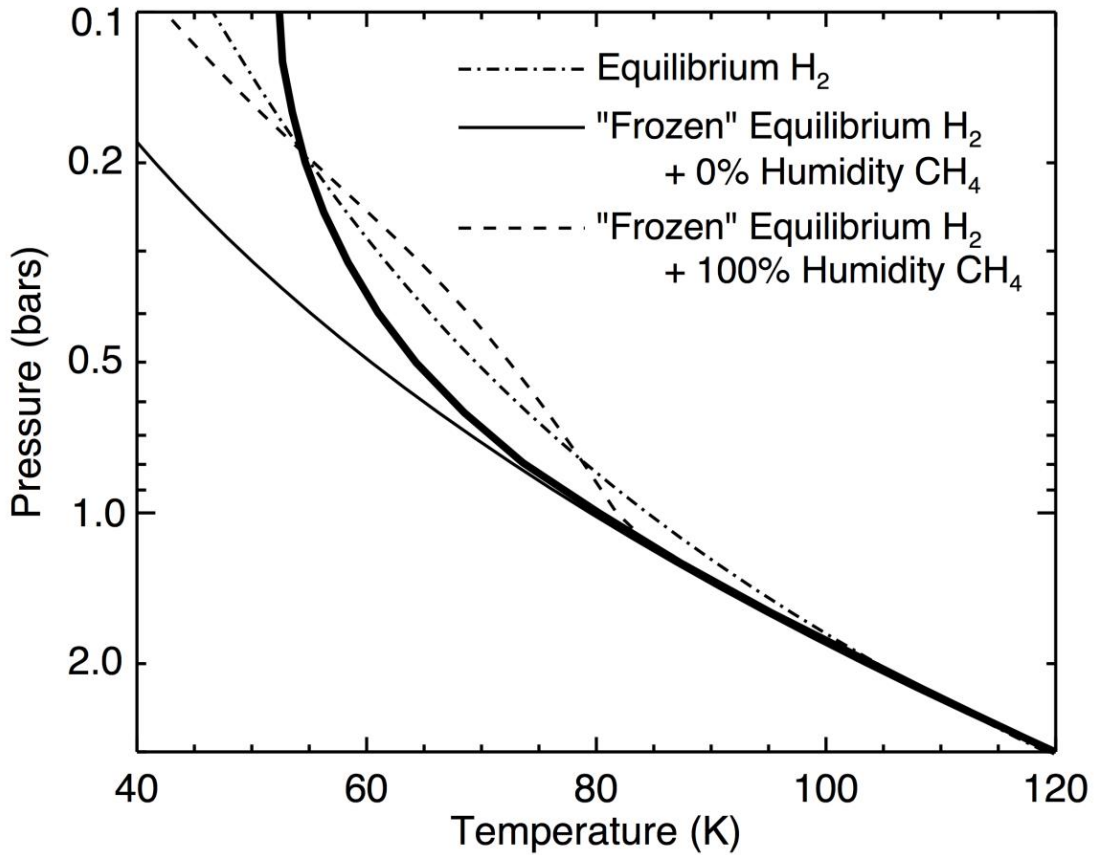

**Figure 16**. Comparison of the derived standard model temperature profile (thick solid curve) and various adiabatic profiles extended upward from the 3-bar pressure level. The profile is inconsistent with an adiabatic profile that conforms to $H_2$ undergoing active equilibration between its *para* and *ortho* states, denoted here as "Equilibrium $H_2$". It is somewhat shallower than a profile with para and ortho states that are in equilibrium at each level but without accounting for latent heat of exchange effects, denoted as "Frozen Equilibrium $H_2$". The profile is inconsistent with adiabatic conditions and the specific heat and latent heat effects of $CH_4$ if fully saturated above the saturation level "100% Humidity $CH_4$", assuming 2.3% $CH_4$ by volume at depth. The disagreement is worse for 4.0% $CH_4$ at depth, consistent with the analysis of radio-occultation results (Lindal et al. 1987, Sromovsky et al. 2011). The retrieved profile could be consistent with an adiabatic profile with a saturated abundance somewhere between 0% and 100% relative humidity.

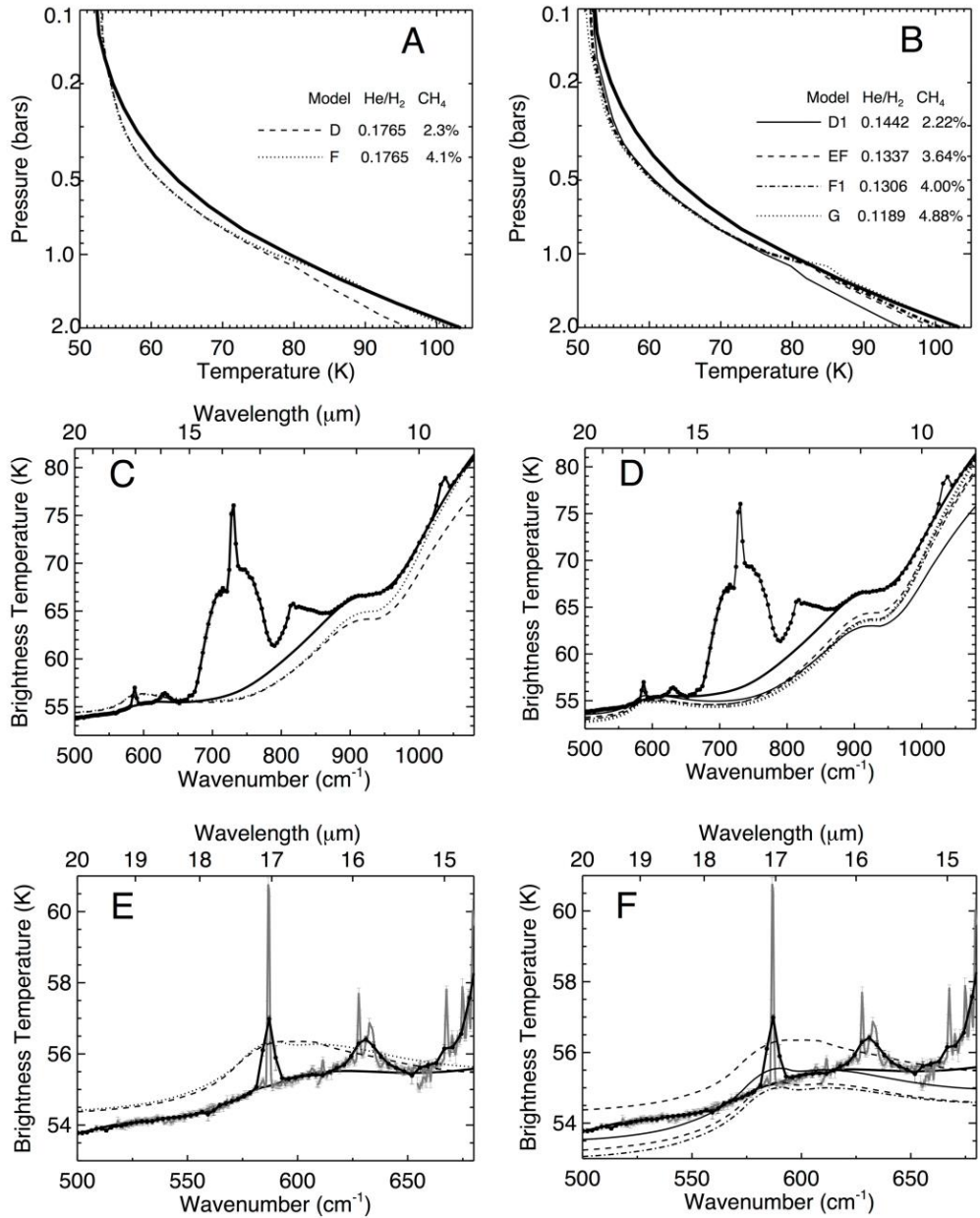

**Figure 17**. Comparisons between our standard model profile and radio occultation experiment profiles. *Panels A, B*. Temperature profiles. *Panels C, and E.*. H$_2$ collision-induced spectra corresponding to the profiles in Panel A. *Panels D and F*. Similar spectra corresponding to the profiles in Panel B. The thick solid line represents the profile of our standard model in Panels A and B and its H$_2$ CIA spectrum in Panels C- F. Panels A, C and E show radio-occultation profiles D and F of Lindal et al. (1987), both of which assume a 15/85 He/H$_2$ ratio with different assumptions for the CH$_4$ mole fraction in the deep atmosphere, and their spectra. Panels B, D and F show radio-occultation profiles D1, EF, F1 and G of Sromovsky et al. (2011), with the legend giving the assumed He/H$_2$ ratio and deep-atmosphere CH$_4$ mole-fraction values for each model, and their spectra. The observed spectra are plotted as in Fig. 11.

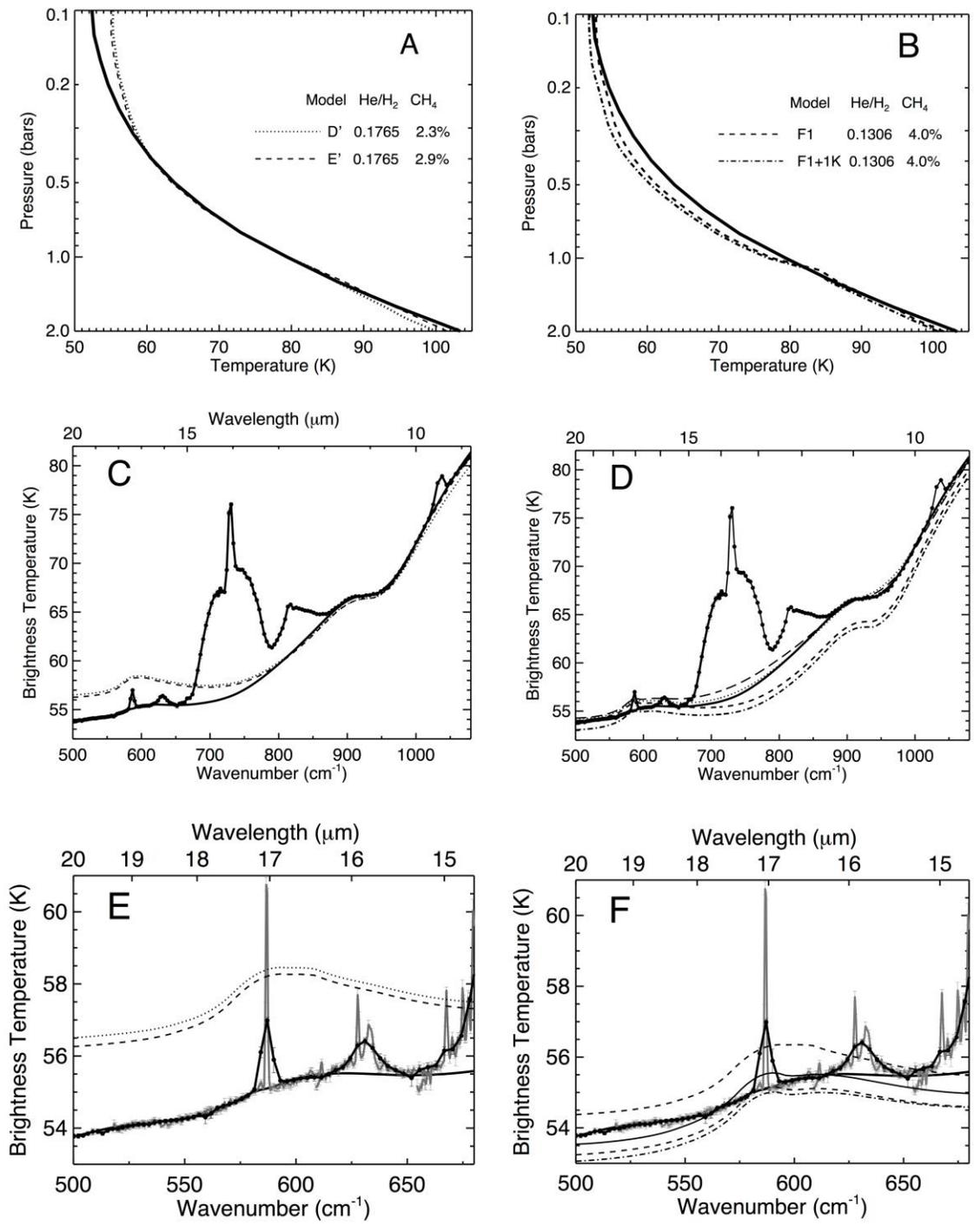

**Figure 18** (previous page) Comparisons between our standard model profile and modified radio occultation experiment profiles. *Panels A, B*. Temperature profiles. *Panels C, D, E, F* Corresponding $H_2$ collision-induced spectra, using the same organization as for Fig. 17. Just as in that figure, the thick solid line represents the profile of our standard model in Panels A and B, and its $H_2$ CIA spectrum in the Panels C-F. Panel A shows the Lindal et al. profile D with temperatures increased by 4% (dotted line) and their profile E with temperature increased by 3.5% (dashed line) with corresponding $H_2$ CIA spectra in Panels C and E. These increased temperatures would result from a corresponding increase of the assumed mean-molecular weight of the atmosphere at each level. Panel B shows the Sromovsky et al. Model F1 temperature profile with a 1-K uniform temperature increase (short-dashed line) and corresponding $H_2$ CIA spectra in Panels D and F. Panels D and F also show two spectra with Model F1+1K and the addition of a particulate monolayer at 31.6 µbar (152.7 K) that has an optical thickness of 5.0 x $10^{-6}$ at 11 µm (909 cm$^{-1}$), with optical thicknesses that are inversely dependent on wavelength (long-dashed line) and the fourth power of wavelength (dotted line). The observed spectra are plotted as in Figures 11 and 17.

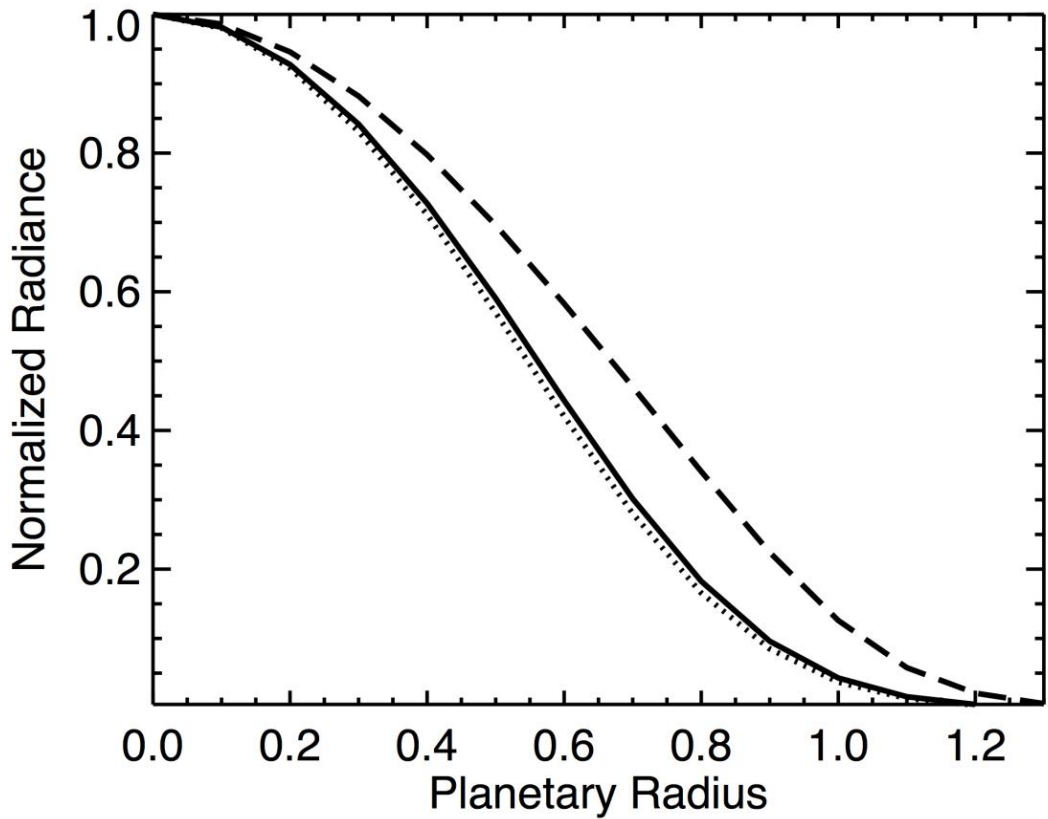

**Figure 19.** Center-to-limb behavior of the radiance from Uranus at 11 μm. The thick solid line corresponds to the response of our standard model. The dotted line corresponds to Model F of Lindal et al. (1987), as shown in **Fig. 17A,** with spectra shown in **Figs. 17C** and **17E**. The dashed line corresponds to Model F' of Sromovsky et al. (2011) that is warmed uniformly by 1 K (**Fig 18B)** and with a particulate opacity layer at 31.6 μbar with an optical thickness (5.0 x $10^{-6}$) that increases the model radiance to the level observed at 11 μm **(Figs. 18D** and **18F)**. The radiances shown have been blurred to correspond to a diffraction-plus-seeing equivalent FWHM of 0.5 arcsec.

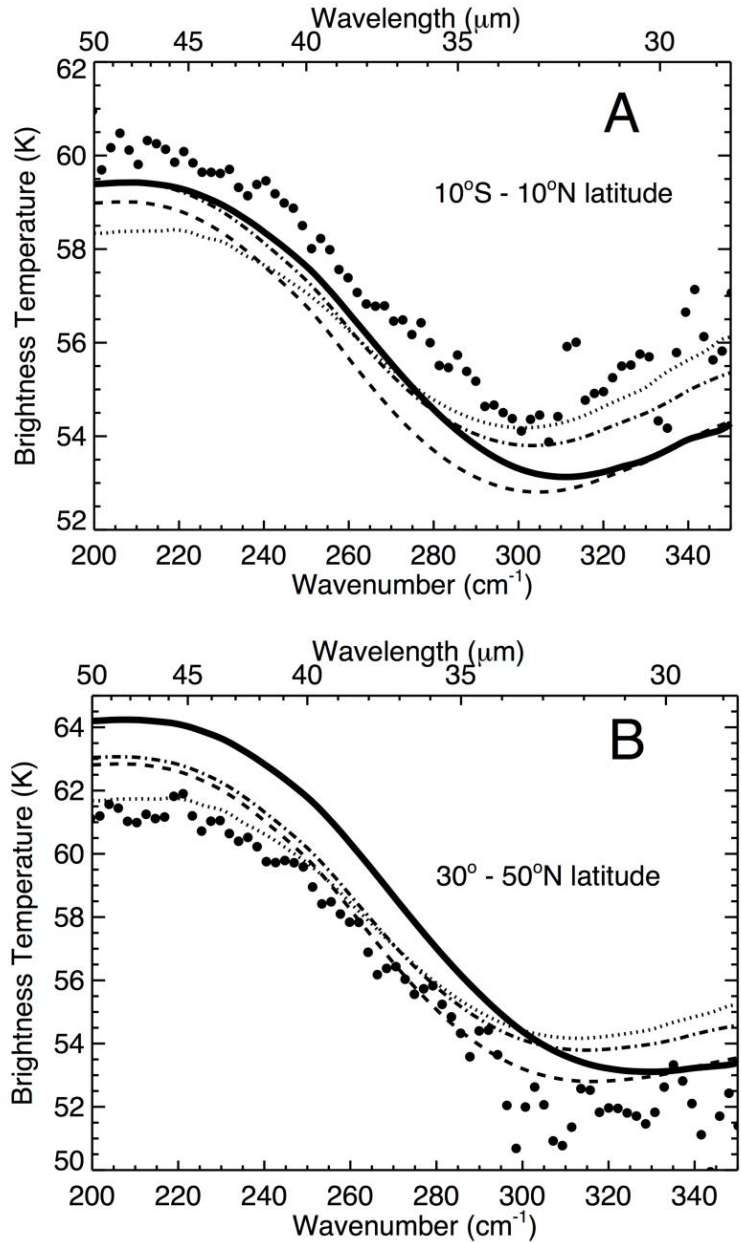

**Figure 20**. Comparison between brightness-temperature spectra predicted by our global-mean model and Voyager-2 IRIS results (e.g. Conrath et al. 1987 and references therein). *Panel A*. Comparison of the spectrum predicted by our global-mean model (solid line) and binned data (filled circles) for the warmest latitudes detected in Uranus by the IRIS experiment for a mean emission angle of 59˚, together with the spectra predicted by Model F of Lindal et al. (1987), given by the dotted line, and by Model F1 of Sromovsky et al (2011), given by the dashed line, the same model with a uniform 1-K temperature increase (i.e. the F1+1K Model in Fig. 18). *Panel B*. Comparison of our global-mean model and binned data for the coldest latitudes detected by IRIS for a mean emission angle of 34˚, together with the spectra predicted by the same models as in Panel A.

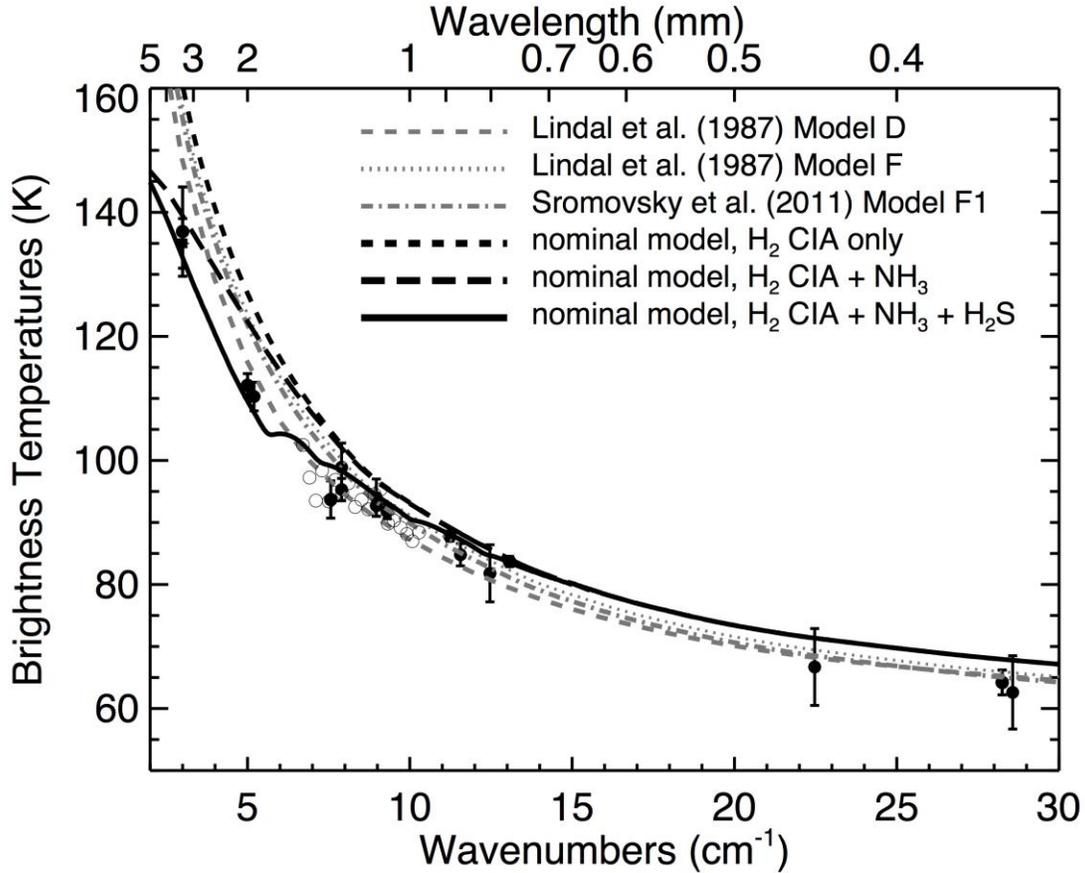

**Figure 21**. Model spectra in the far-infrared and submillimeter region compared with observations. Filled circles represent observations summarized by Griffin and Orton (1993); open circles represent the spectrum of Serabyn and Weisstein (1996). Various spectral models are shown with lines as labeled in the key, which replicates their use in **Fig. 17**. Model EF of Sromovsky et al. (2011) produces nearly identical spectra in this region to their Model F1, which is shown. The spectrum labeled "nominal model" corresponds to the nominal temperature structure developed in this paper (**Fig. 16**). The spectra shown in this figure for the Lindal et al. (1987) and Sromovsky et al. (2011) models assume only $H_2$ collision-induced absorption, as well as the nominal model identified by the long-dashed line. The "$H_2$ CIA + $NH_3$" model adds the opacity of $NH_3$ inversion and rotational lines that improves the fit to the microwave spectrum. The "$H_2$ CIA + $NH_3$ + $H_2S$" line shows a modification that substantially improves the overall fit by adding absorption by $H_2S$ rotational lines. Details of the abundance and vertical distribution of $NH_3$ and $H_2S$ are given in the text.